\documentclass[aps,prd,twocolumn,showpacs,superscriptaddress,groupedaddress]{revtex4}  
\usepackage{graphicx}  
\usepackage{dcolumn}   
\usepackage{bm}        
\usepackage{amssymb}   
\usepackage{xcolor}
\usepackage{hyperref}

\begin{document}

\widetext

\newcommand{\beq}{\begin{eqnarray}}
\newcommand{\eeq}{\end{eqnarray}}
\newcommand{\nn}{ \nonumber}
\newcommand{\half}{\frac{1}{2}}
\newcommand{\chib}{{\bar \chi}}
\newcommand{\qb}{{\bar q}}
\newcommand{\Db}{{\bar {\cal D}}}
\newcommand{\D}{{\cal D}}
\newcommand{\ben}{\begin{enumerate}}
\newcommand{\een}{\end{enumerate}}
\newcommand{\bpsi}{{\bar \psi}}
\newcommand{\Bpsi}{{\bar \Psi}}
\newcommand{\etab}{{\bar \eta}}
\newcommand{\vev}[1]{{\langle #1 \rangle}}
\newcommand{\Dslash}{{\not \hspace{-4pt} D} }
\newcommand{\Dslashexp}{{\not \hspace{-1pt} D}}
\newcommand{\partialslash}{{\not \hspace{-4pt} \partial}}
\newcommand{\Aslash}{{\not \hspace{-4pt} A}}
\newcommand{\Pislash}{{\not \hspace{-4pt} \Pi}}
\newcommand{\Lcal}{{\cal L}}
\newcommand{\nnn}{ \nonumber \\ }
\newcommand{\ddd}{\nnn &&}
\newcommand{\angstrom}{\mbox{\normalfont\AA}}
\newcommand{\tr}{\mathop{{\hbox{Tr} \, }}\nolimits}
\def\etal{{\it et al.}}

\author{Carleton DeTar}
\author{Christopher Winterowd}
\affiliation{Department of Physics and Astronomy \\ University of Utah, Salt Lake City, Utah, USA}
\author{Savvas Zafeiropoulos}
\affiliation{Institut f{\"u}r Theoretische Physik - Johann Wolfgang Goethe-Universit{\"a}t \\ Max-von-Laue-Str. 1, 60438 Frankfurt am Main, Germany}
\title{Lattice Field Theory Study of Magnetic Catalysis in Graphene}
\date{\today}

\begin{abstract}
We discuss the simulation of the low-energy effective field theory (EFT) for graphene in the presence of an external magnetic field. Our fully nonperturbative calculation 
uses methods of lattice gauge theory to study the theory using a hybrid Monte Carlo approach. We investigate the phenomenon of magnetic catalysis in the context of graphene by studying
the chiral condensate which is the order parameter characterizing the spontaneous breaking of chiral symmetry. In the EFT, the symmetry breaking pattern is given by $U(4) \to U(2) \times U(2)$. We also comment on the difficulty, in this lattice formalism, of studying the time-reversal-odd condensate characterizing the ground state in the presence of a magnetic field.
Finally, we study the mass spectrum of the theory, in particular the Nambu-Goldstone (NG) mode as well as the Dirac quasiparticle, which is predicted to obtain a dynamical mass. 
\end{abstract}

\maketitle

\section{\label{sec:Intro}Introduction}
The discovery of graphene counts as one of the most important developments in recent years \cite{Novoselov}. The study of graphene has seen tremendous growth,
in part due to its novel many-body and electronic properties \cite{CastroNeto}. In particular, there have been many studies of graphene
in the presence of an external magnetic field \cite{Goerbig}.

Graphene, a two-dimensional hexagonal lattice of Carbon atoms, has an unusual band structure consisting of a conduction and a valence band that come together at two inequivalent corners 
of the Brillouin zone. At these valleys, or so-called ``Dirac points'', the two bands take the shape of cones. The consequence of this band structure is that the low energy excitations are described by massless Dirac fermions, which have linear dispersion.
These Dirac fermions have a Fermi velocity that satisfies $v_F/c \approx 1/300$, and thus one does not have Lorentz invariance. Furthermore, this small Fermi velocity renders the interaction between the Dirac quasiparticles 
to be essentially Coulombic. 

The effective coupling of the theory is particularly large, $\alpha_g \equiv 2e^2/((\epsilon+1) v_F 4\pi) > 1$, and thus this is a strong-coupling problem. Here, the strength of the interaction between the quasiparticles is controled by the dielectric constant, $\epsilon$, of the substrate on which the graphene layer sits. For suspended graphene $\epsilon = 1$, while for graphene on a silicon oxide substrate $\epsilon_{\rm Si0_2} \approx 3.9$. It is known that for sufficiently large coupling, the EFT undergoes a phase transition, as has been shown in 
several previous studies \cite{Drut1, Drut2, Hands1, Hands2}. This transition is characterized by the appearance of a nonzero value of the condensate, $\vev{\Bpsi \Psi}$, as well as a dynamical mass
for the fermion.

In the presence of an external magnetic field, many field theories are thought to undergo spontaneous symmetry breaking. This phenomenon is known as magnetic catalysis and is hypothesized to be universal. In the context of the graphene EFT, the magnetic
catalysis scenario leads to the appearance of a dynamically generated Dirac mass and is proposed to account for several of the quantum-Hall plateaus that appear at large values of the magnetic field \cite{ZhangQHE,JiangQHE, Kennett}.

Magnetic catalysis has also been the focus of recent interest in lattice quantum chromodynamics (LQCD). Several studies have shown that the chiral condensate increases with magnetic field at zero temperature ($m_{\pi} = 200-400 ~\text{MeV}$, $eB$ up to $1 ~\text{GeV}^2$) \cite{Buividovich,Braguta,Cohen}. This confirms the idea that the background magnetic field aids chiral symmetry breaking in the vacuum.
However, it was also discovered that with $T \geq 140 ~\text{MeV}$ and physical pions, the opposite occurs and the chiral condensate decreases with the external magnetic field \cite{Bali1,Bali2}.

The remainder of this article is organized as follows. In section \ref{sec:GrapheneEFT} we discuss the continuum EFT, its symmetries, as well as our lattice setup.
In section \ref{sec:Catalysis} we discuss the physical mechanism behind magnetic catalysis and review the results that apply to graphene. In section \ref{sec:Simulation} we review the methods used to sample the Feynman path integral numerically. Section \ref{sec:Observables} introduces the various observables that we use to study the graphene EFT in the presence of an external magnetic field. Section \ref{sec:Results} contains our results for the various condensates, and the mass spectrum of the EFT.
Finally, in Section \ref{sec:Conclusion}, we conclude and discuss the interpretation of these results. Some preliminary results of this article first appeared in Ref.~\cite{DPF2015} and Ref.~\cite{GrapheneLetter}.
\section{\label{sec:GrapheneEFT}Graphene EFT}
\subsection{\label{sec:ContEFT}Continuum}
The continuum EFT, describing the low-energy properties of monolayer graphene, contains two species of four-component Dirac spinors in $(2+1)$ dimensions interacting via a Coulomb interaction \cite{DrutSon}.
The counting of the fermionic degrees of freedom is as follows: $2$ sublattices $\times~2$ Dirac points $\times~2$ spin projections of the electrons. The sublattice degree of freedom appears due to the fact that the hexagonal lattice is bipartite, consisting of two inequivalent triangular sublattices.
The interaction is introduced via a scalar potential which lives in $(3+1)$ dimensions. The continuum Euclidean action is given by 
\beq
\label{ContinuumEFT}
\nn
 S_E &=& \int dt d^2 x \sum_{\sigma=1,2} \Bpsi_{\sigma} \Dslash[A_0] \Psi_{\sigma} \\ &+& \frac{(\epsilon + 1)}{4e^2} \int dt d^3 x ~ (\partial_i A_0)^2.
\eeq
The Dirac operator is given by 
\beq
\Dslash[A_0] = \gamma_0 \left( \partial_0 + i A_0 \right) + v_F \sum_{i=1,2} \gamma_i \partial_i.
\eeq
The basis used for the four-component Dirac spinors is as follows
\beq
\label{DiracSpinorBasis}
\Psi^{\top}_{\sigma} = \left( \psi_{K_+ A \sigma}, \psi_{K_+ B \sigma}, \psi_{K_- B \sigma}, \psi_{K_- A \sigma}\right),
\eeq
where $K_+, K_-$ refer to the Dirac points, $\sigma$ refers to the electron's spin, and $A, B$ refer to the sublattices.
Here we use a reducible representation of the gamma matrices in $(2+1)$ dimensions, constructed from the two inequivalent irreducible representations, which is given by
\beq
\label{Gammas1}
\gamma_{\mu} = \left(\begin{array}{cc} \sigma_{\mu} & 0 \\ 0 & -\sigma_{\mu} \end{array}\right), ~~~ \mu = 0, 1, 2,
\eeq
where $\sigma_0 \equiv \sigma_3$. In $(3+1)$ dimensions, a similarity transformation, $S^{\dagger} \gamma_{\mu} S = - \gamma_{\mu}$, relates the two representations, with $S = \gamma_5$.
In $(2+1)$ dimensions, $\gamma_5$ does not exist as $\displaystyle \prod^{2}_{\mu=0} \sigma_{\mu} \propto  \bm 1$. One can also define additional matrices which generate symmetry transformations in the graphene EFT
\beq
\label{Gammas2}
&&\tilde{\gamma}_{4} = \left(\begin{array}{cc} 0 & \bm 1 \\ \bm 1 & 0 \end{array}\right), ~~~ \tilde{\gamma}_{5} = \left(\begin{array}{cc} 0 & \bm 1 \\ -\bm 1 & 0 \end{array}\right), \\
\ddd \tilde{\gamma}_{4,5} \equiv -\tilde{\gamma}_{4} \tilde{\gamma}_{5} = \left(\begin{array}{cc} \bm 1 & 0 \\ 0 & -\bm 1 \end{array}\right),
\eeq
where $\{ \tilde{\gamma}_4, \gamma_{\mu} \} = \{ \tilde{\gamma}_5, \gamma_{\mu} \} = \{ \tilde{\gamma}_4, \tilde{\gamma}_5 \} = 0$.

The fermionic part of the action in (\ref{ContinuumEFT}) has a global $U(4)$ symmetry whose generators are given by 
\beq
\label{unbrokenGens}
&& \bm 1 \otimes \bm 1, \quad \bm 1 \otimes \sigma_{\mu}, \quad \tilde{\gamma}_{4,5} \otimes \bm 1, \quad \tilde{\gamma}_{4,5} \otimes \sigma_{\mu} \\
\label{brokenGens}
&& \tilde{\gamma}_4 \otimes \bm 1, \quad \tilde{\gamma}_4 \otimes \sigma_{\mu}, \quad i\tilde{\gamma}_5 \otimes \bm 1, \quad i\tilde{\gamma}_5 \otimes \sigma_{\mu}.
\eeq
These generators have the four-dimensional sublattice-valley subspace tensored with the two-dimensional spin subspace.
The appearance of a Dirac mass term, of the form $\displaystyle m \sum_{\sigma=1,2} \Bpsi_{\sigma} \Psi_{\sigma}$ breaks the $U(4)$ symmetry down to \\ $U(2) \times U(2)$, whose generators are given by (\ref{unbrokenGens}).
Thus, the formation of a nonzero value of the condensate $\vev{\Bpsi_{\sigma} \Psi_{\sigma}}$, would signal spontaneous symmetry breaking and the appearance of eight Nambu-Goldstone (NG) bosons. In
Appendix \ref{sec:FermionAppendix}, we provide the details of the fermion bilinears relevant to this study.

As we have mentioned, the action in (\ref{ContinuumEFT}) is diagonal in spin space and has a $U(4)$ symmetry described by the generators in (\ref{unbrokenGens}) and (\ref{brokenGens}).  
Although our lattice simulations do not take this into account, in real graphene this $U(4)$ symmetry is explicitly broken in the presence of an external magnetic field by the Zeeman term in the Hamiltonian
\beq
\label{ZeemanHamiltonian}
\mathcal{H}_Z = -\mu_B B\int dt d^2x~ \Psi^{\dagger} \sigma_3 \Psi,
\eeq
where $\mu_B$ is the Bohr magneton and $\sigma_3$ acts in spin space. Although we are dealing with Dirac fermions, this term is needed as the four-dimensional spinor structure is constructed from the sublattice and valley degrees of freedom. In the graphene EFT, the orbital and pseudospin quantum numbers determine the Landau-level index, while the electron's spin degree of freedom couples to the magnetic field in a nonrelativistic way. In contrast, for relativistic Dirac fermions coupled to an external magnetic field, the spin degree of freedom in the direction of the field combines with the orbital quantum number to determine the Landau-level index. 
The Zeeman term explicitly breaks the $U(4)$ symmetry down to $U(2)_{\uparrow} \times U(2)_{\downarrow}$ with the generators of $U(2)_{\sigma}$ given by
\beq
\label{U2Generators}
\bm 1 \otimes P_{\sigma}, \quad \tilde{\gamma}_4 \otimes P_{\sigma}, \quad i \tilde{\gamma}_5 \otimes P_{\sigma}, \quad \tilde{\gamma}_{4,5} \otimes P_{\sigma},
\eeq
where we have introduced the spin projection operator 
\beq
P_{\sigma} \equiv \frac{1}{2}( 1 \pm \sigma_3 ).
\eeq

Although the Zeeman term lifts the spin degeneracy of each Landau level, the size of this perturbation is very small even in the presence of large magnetic fields. This can be seen by considering 
\beq
\epsilon_{LL} &\equiv& \sqrt{\hbar v^2_F |eB|/c} = 26 \sqrt{B[\text{T}]}~\text{meV}, \\ 
\epsilon_Z &\equiv & \mu_B B = 5.8\times 10^{-2} B[\text{T}]~\text{meV},
\eeq
where $\epsilon_{LL}$ is the energy separating the lowest Landau level ($n=0$) from its neighbor ($n=1$), and $\epsilon_Z$ is the Zeeman energy. For even the largest magnetic fields available in the laboratory ($B \sim 50~\text{T}$), $\epsilon_Z$ is only a small fraction of $\epsilon_{LL}$.

 Apart from the long-range Coulomb interaction present in (\ref{ContinuumEFT}), in the complete lattice
theory there are numerous short-range electron-electron interactions. These lattice-scale interaction terms are allowed by the point-group symmetry of the underlying hexagonal lattice, $C_{6v}$ \cite{Aleiner}.
As a result, these terms break the much larger $U(4)$ symmetry present in our continuum EFT. The couplings associated with these terms can vary in sign and are strongly renormalized at energies on the order of the bandwidth, $v_F/a$, where $a$ is the spacing of the hexagonal lattice \cite{Kharitonov}.
Taking these renormalized couplings into account can have a decisive effect on the selection of the ground state in the full theory. In our lattice gauge model, we consider only a long-range Coulomb interaction mediated by a scalar potential and ignore the Zeeman interaction as well as 
other lattice-scale interactions. 
\subsection{\label{sec:Lattice}Lattice}
Taking the continuum EFT (\ref{ContinuumEFT}), which is valid up to some cutoff $\Lambda$, we now discretize it on a cubic lattice. It is important to emphasize that this lattice is not 
directly related to the original honeycomb lattice of graphene. Lattice methods in gauge theories have shown to be useful in elucidating the nonperturbative aspects of strongly interacting theories such as quantum chromodynamics (QCD) \cite{DeGrandDeTar}.
Thus, we chose to apply these methods to the study of the graphene EFT.

We represent the gauge potential, $A_0(n)$, via the variable $U_0(n) \equiv \exp\left(i a_tA_0(n)\right)$, which lives on the temporal links of the lattice. We take the fermion fields to live on the sites of the lattice. 
To avoid complications arising from an unwanted bulk phase transition \cite{KogutStrouthos}, we use the so-called noncompact $U(1)$ gauge action 
\beq
\label{NCGaugeAction1}
S^{(NC)}_G = a^3_s a_t \frac{\beta}{2} \sum_n \sum^{3}_{i=1} \frac{1}{a^2_s}\left(A_0(n) - A_0(n+\hat{i})\right)^2,
\eeq
where $\beta =(\epsilon+1)/2e^2$ and $a_s$ and $a_t$ are the lattice spacings in the spatial and temporal directions respectively. Here $n=(n_0,n_1,n_2,n_3)$ labels the lattice site and $\hat{i}$ is the unit vector in the ith direction. Using the dimensionless field $\hat{A_0}(n) = a_t A_0(n)$, we write the above action as
\beq
\label{NCGaugeAction2}
S^{(NC)}_G = \xi \frac{\beta}{2} \sum_n \sum^{3}_{i=1} \left(\hat{A_0}(n) - \hat{A_0}(n+\hat{i})\right)^2,
\eeq
where $\xi \equiv a_s/a_t$ is the anisotropy parameter, controlling the ratio of the spatial lattice spacing to the temporal lattice spacing.

The discretization of the fermions has well-known complications encoded in the Nielsen-Ninomiya no-go theorem \cite{NielsenNinomiya}. Namely, one wants to remove unphysical ``doubler'' modes while still preserving some part of the continuum $U(4)$
symmetry. The method we use in this study employs the staggered-fermion formulation \cite{KogutSusskind}, which eliminates some of the doublers while preserving a remnant of the $U(4)$ symmetry. The staggered-fermion action reads
\beq
\label{GrapheneFermLattice}
\nn
S_F &=& a^2_s a_t \sum_{n} \bigg[ \frac{1}{2a_t} \chib_n \left(U_0(n)\chi_{n + \hat{0}} - U^{\dagger}_0(n-\hat{0})\chi_{n - \hat{0}}\right) \\ \nn &+&  
\frac{1}{2a_s}v_F\sum_{i=1,2} \eta_{i}(n) \chib_n \left(\chi_{n + \hat{i}} - \chi_{n - \hat{i}}\right) \\ &+& m\chib_n\chi_n \bigg],
\eeq
where the fermions live in $(2+1)$ dimensions and $\eta_1(n) = (-1)^{n_0}$, $\eta_2(n) = (-1)^{n_0 + n_1}$ are the Kawamoto-Smit phases which appear due to spin diagonalization
of the naive fermion action. The fields $\chi$ and $\chib$ are one-component Grassmann variables. We have introduced a mass term which explicitly breaks the remnant chiral symmetry but is needed in our simulations as an infrared regulator and to investigate SSB. We will later take the infinite-volume limit followed by the chiral limit, $m \to 0$.
We can also write the fermion action in terms of the dimensionless lattice quantities $\hat{\chib}_n = a_s \chib_n$, $\hat{\chi}_n = a_s \chi_n$, $\hat{m} = a_t m$
\beq
\label{GrapheneFermLatticeDimensionless}
\nn
S_F &=& \sum_{n} \bigg[ \frac{1}{2} \hat{\chib}_n \left(U_0(n)\hat{\chi}_{n + \hat{0}} - U^{\dagger}_0(n-\hat{0})\hat{\chi}_{n - \hat{0}}\right) \\ \nn &+& 
\frac{v_F}{2\xi}\sum_{i=1,2} \eta_{i}(n) \hat{\chib}_n \left(\hat{\chi}_{n + \hat{i}} - \hat{\chi}_{n - \hat{i}}\right) \\ &+& \hat{m}\hat{\chib}_n\hat{\chi}_n \bigg].
\eeq
In $(2+1)$ dimensions, each species of staggered fermions represents two identical four-component Dirac spinors. In lattice QCD simulations, this degree of freedom, commonly referred to as ``taste'',
is unwanted, and considerable effort goes into eliminating it. In our case, the taste degree of freedom is desirable, as we are attempting to simulate two identical species of Dirac fermions representing the low energy excitations of graphene.
The taste degree of freedom becomes more apparent once one performs a change of basis, the details of which are relegated to Appendix \ref{sec:SpinTasteAppendix}. At zero mass, the lattice action in (\ref{GrapheneFermLatticeDimensionless}) retains a $U(1) \times U(1)_{\epsilon}$ remnant of the continuum $U(4)$ symmetry.
The $U(1)$ symmetry corresponds to fermion number conservation and is given by
\beq
\label{U1Staggered}
\chi(x) &\to& \exp \left(i\alpha \right) \chi(x), \\ \nn \chib(x) &\to& \chib(x) \exp \left(-i \alpha \right).
\eeq
The $U(1)_{\epsilon}$, or ``even-odd'' symmetry, is given by 
\beq
\chi(x) &\to& \exp \left(i\beta \epsilon(x) \right) \chi(x), \\ \nn \chib(x) &\to& \chib(x) \exp \left(i \beta \epsilon(x) \right),
\eeq
where $\epsilon(x) \equiv \left( -1 \right)^{x_0 + x_1 + x_2}$. The $U(1)_{\epsilon}$ symmetry is spontaneously broken with the appearance of a nonzero value of the condensate, $\vev{\chib \chi}$. Thus, on the lattice, one expects there to be a single NG boson as a result of spontaneous symmetry breaking. For the study of the time-reversal-odd condensate, the calculation is a bit more intricate with staggered fermions. Namely, in order to compare with continuum results, one is interested in distinguishing the spin projections of the condensate, which in staggered language, corresponds to the taste degree of freedom. Due to taste-breaking terms in the staggered action at $O(a)$ (see Appendix \ref{sec:SpinTasteAppendix}), this becomes impossible as these terms mix the various tastes, or in graphene language, spin projections. 

In this study we use the improved staggered fermion action, asqtad \cite{Orginos}. We outline the various improvements in the following paragraphs.

Although staggered fermions have some advantages, being cost effective in numerical simulations and retaining a remnant chiral symmetry, it is known that at finite lattice spacing the violations
of taste symmetry can be significant. In the case of QCD, for example, simulations with unimproved staggered fermions on fine lattices ($a=0.05$ fm) obtain splittings within the pion taste-multiplet ($O(100~\text{MeV})$)
which are of the order of the physical pion mass \cite{MILCStaggeredReview}. It was found that taste violations are a result of the exchange of high-momentum gluons \cite{LagaeSinclair}.
These couplings are eliminated by a process of link ``fattening'', which replaces the links $U_{\mu}(n)$ with a combination of paths that connect the site $n$ with $n+\hat{\mu}$. This is done in the asqtad action.

Another improvement which has proven useful in lattice QCD 
simulations is the so called tadpole improvement \cite{LepageMackenzie}. This, too, is used in the asqtad action. This improvement is motivated by the observation that new vertices, with no continuum analog, appear in lattice perturbation theory. These vertices are suppressed by powers of the lattice spacing but
can lead to UV-divergent diagrams which cancel the lattice spacing dependence of the vertex. In the graphene EFT, the tadpole factor, $u_0$, is defined in the following way
\beq
u_0 = \left( \vev{U^{(p)}} \right)^{1/2},
\eeq
where $\vev{U^{(p)}}$ is the volume-averaged expectation value of the space-time-oriented plaquette. Tadpole improvement consists of dividing each link in the temporal direction by $u_0$. Thus, each fat link
receives a factor $1/u^{l_t}_0$, where $l_t$ is the number of steps the path takes in the temporal direction.

One can further improve the staggered-fermion action by introducing a third-nearest-neighbor hopping. This is known as the Naik term \cite{Naik}. With an appropriate weighting of the nearest neighbor and the third-nearest neighbor terms,
one can obtain an improved free dispersion relation. Thus, using all of the above-mentioned ingredients, one has a tadpole-improved action which is free of discretization errors up to $O(a^2)$. 
This is what is known as the improved asqtad action \cite{Orginos}. 

Although previous lattice studies have used various levels of improvement for staggered fermions \cite{Giedt,Drut3}, further reducing taste violations by using the asqtad action should 
come closer to realizing the continuum theory. 

The introduction of an external magnetic field on the lattice proceeds in the following manner \cite{WieseAlHashimi}. In the continuum, a homogenous magnetic field perpendicular to the sheet of graphene can be described by the Landau-gauge vector potential, $A_{\mu} = \delta_{\mu,2}Bx_1$.
The spatial links are similarily modified, with a special prescription at the boundary of the lattice due to the periodic boundary conditions satisfied by the gauge links
\beq
\label{ExtMagFieldLinks}
U_y(n) &=& e^{ia^2_seB n_x}, \\
U_x(n) &=& \left\{ \begin{array}{ll} 1 &, n_x \neq N_s-1 \\ 
                    e^{-ia^2_seBN_x n_y} &, n_x = N_s -1
                   \end{array} \right. .
\eeq
where $N_s = N_x = N_y$, and $n_x, n_y = 0, 1, \dots , N_s-1$.
Furthermore, the toroidal geometry demands that the magnetic flux through the lattice be quantized as follows 
\beq
\label{FluxQuantization}
\Phi_B \equiv \frac{eB}{2\pi} = \frac{N_{B}}{L^2_s},
\eeq
where $L_s = N_s a_s$, is the lattice extent in the spatial direction and $N_B$ is an integer in the range 
\beq
0 \leq N_B \leq \frac{N^2_s}{4}.
\eeq
We note that the spatial links, which describe the external magnetic field, are static and are not updated during the sampling of the path integral.
\section{\label{sec:Catalysis}Magnetic Catalysis}
The phenomenon of magnetic catalysis is a fascinating example of dynamical symmetry breaking \cite{Miransky1,Miransky2,Miransky3,Miransky4}. In this scenario, an external magnetic field acts as a catalyst for fermion-antifermion pairing, even if they are
weakly coupled. First studied in the context of the Nambu-Jona-Lasinio model and quantum electrodynamics in both $(2+1)$ and $(3+1)$ dimensions, magnetic catalysis has been predicted to occur
also for planar condensed-matter systems, including graphene \cite{Khveshchenko,MiranskyGraphene1,MiranskyGraphene2,MiranskyGraphene3}. Although various perturbative approaches have been used to study magnetic catalysis in various settings, it is useful to apply lattice methods, as one is able to 
take a fully nonperturbative approach.

One can begin to understand the mechanism responsible for magnetic catalysis by first considering free Dirac fermions in the presence of an external magnetic field. This mechanism involves a dimensional 
reduction, $D \to D-2$, due to the magnetic field \cite{Shovkovy}. The Dirac equation in $(2+1)$ dimensions is given by 
\beq
\label{DiracEqn}
(i\gamma^{\mu}\D_{\mu} - m)\Psi = 0,
\eeq
where $\D_{\mu} = \partial_{\mu} - ieA_{\mu}$.
Solving (\ref{DiracEqn}), one finds that the energy levels are given by 
\beq
\label{DiracEnergyLevels}
E_n = \pm \sqrt{2|eB|n + m^2},
\eeq
where $n=0,1,2,\cdots$ is the Landau-level index and is given by a combination of orbital and spin contributions, $n \equiv k + s_z + 1/2$. The Landau levels 
are highly degenerate, with a degeneracy per unit area of $|eB|/2\pi$ for $n=0$, and $|eB|/\pi$ for $n>0$. When the Dirac mass is much smaller than the magnetic scale, $m \ll \sqrt{|eB|}$, the low energy sector 
is completely dominated by the lowest Landau level. Furthermore, the levels at $E_0 = \pm m$ do not disperse which confirms the kinematic aspect of the reduction, $(2+1) \to (0+1)$, due to the presence of the magnetic field.

An interesting consequence of an external field on Dirac fermions in $(2+1)$ dimensions is the appearance of a nonzero value for the condensate, $\vev{\Bpsi \Psi}$. This supports the existence of spontaneous symmetry breaking.
One can calculate the free fermion propagator in the presence of an external field \cite{Schwinger}, and use the result to arrive at \cite{DittrichGies, DittrichReuter}
\beq
\label{PBP2+1} \nn
\vev{\Bpsi \Psi} &=& -\frac{1}{2\pi}\biggl(m \sqrt{2eB} \zeta \left(\frac{1}{2}, 1+\frac{m^2}{2eB}\right) \\ &+& eB - 2m^2 \biggr),
\eeq
where $\zeta(s,a)$ is the Hurwitz zeta function. Taking the limit $m \to 0$ of (\ref{PBP2+1}), one obtains
\beq
\label{PBP2+1Final}
\lim_{m \to 0_+} \vev{\Bpsi \Psi}(B,m) = -\frac{eB}{2\pi}.
\eeq
Of course, by introducing interactions among the fermions, this result will be modified. Calculations using Schwinger-Dyson equations predict the dynamical mass of the fermion to satisfy \cite{Shovkovy}
\beq
\label{DynamicalMass}
m_{\rm dyn} \propto \alpha_g \sqrt{eB}.
\eeq


In addition to the condensate that one normally considers when discussing spontaneous chiral symmetry breaking, there are other condensates which can describe the ground state of the graphene EFT
in the presence of a magnetic field. In our study we also consider a time-reversal-odd condensate, $\Delta_H \equiv \vev{ \Bpsi \tilde{\gamma}_{4,5} \Psi}$. While the former condensate leads to a Dirac
mass in the graphene EFT, the latter condensate gives rise to a Haldane mass \cite{Haldane}. The ground state at filling factor $\nu=0$, corresponding to a half-filled lowest Landau level, is posited to support a nonzero value for both condensates in the absence of Zeeman splitting. In this work,
we study the flux dependence of both condensates in the chiral and zero-temperature limits.
\section{\label{sec:Simulation}Simulation Details}
We generated our gauge configurations using a $U(1)$ variant of the $\Phi$ algorithm \cite{PhiAlgorithm}. This algorithm falls under a broad class of algorithms collectively known as
hybrid Monte Carlo (HMC) \cite{KogutDuane}. This procedure introduces a fictitious time variable, $\tau$, and a real-valued canonical momentum, $\pi_n$, which is conjugate to the gauge potential $A_0(n)$. One is then able to construct a molecular-dynamics Hamiltonian
\beq
\label{HMCHamiltonian}
\mathcal{H}^{(MD)} = \frac{1}{2} \sum_n \pi^2(n) + S_E,
\eeq
where the momentum term is a Gaussian weight which factors out during the calculation of expectation values, and $S_E$ is our Euclidean lattice action. The equations of motion are then 
\beq
\label{ThetaEOM}
\dot{A}_0(n) &\equiv& \frac{dA_0(n)}{d\tau} = \pi(n), \\
\label{PiEOM}
\dot{\pi}(n) &\equiv& \frac{d\pi(n)}{d\tau} = - \frac{\partial S_E}{\partial A_0(n)}.
\eeq
Integrating these equations over a time interval of length $T$, one generates a new configuration which one either accepts or rejects based on a Monte-Carlo Metropolis decision.
In this study we have used the second-order leapfrog method in integrating (\ref{ThetaEOM}) and (\ref{PiEOM}).

The main question one has when selecting an algorithm to simulate dynamical fermions is how best to deal with the fermion determinant. In our case, we are interested in simulating two continuum species, and thus, for staggered fermions,
one can take advantage of the even-odd symmetry of the staggered Dirac operator. Namely, one first introduces a complex pseudofermion field $\phi$ which lives at each site of the lattice. Based on the identity 
\beq
\label{PFGaussIdentity}
\det(M^{\dagger} M) = \int \mathcal{D} \phi^{\dagger} \mathcal{D} \phi e^{-\phi^{\dagger} \left( M^{\dagger}M \right)^{-1} \phi },
\eeq
one can use the pseudofermions to replace the fermion action by making the identification $M \equiv \Dslash_{st} + m$, where the unimproved staggered Dirac operator is given by
\beq
\label{StaggeredDiracOperator} \nn
(\Dslash_{st})_{x,y} &=& \frac{1}{2} \sum_{\mu} \eta_{\mu}(x) ( U_{\mu}(x)\delta_{y,x+\hat{\mu}} \\ &-& U^{\dagger}_{\mu}(x-\hat{\mu})\delta_{y,x-\hat{\mu}} ).
\eeq
Noting that $M^{\dagger}M$ decouples even and odd sites, the determinant on the LHS of (\ref{PFGaussIdentity}) overcounts the number of degrees of freedom by a factor of two.
This can be corrected by restricting the pseudofermions to the even sites, for example. 

The $\Phi$ algorithm starts by generating a random complex field distributed according to (\ref{PFGaussIdentity}) as well as a Gaussian-distributed momentum. One can then integrate the equations of motion over a trajectory. Specifically, the difficult part in doing this is computing the ``force'' given by (\ref{PiEOM}).
More specifically, we can write this expression as 
\beq
\dot{\pi}_n = -\frac{\partial S_G}{\partial A_0(n)} - \frac{\partial S_F}{\partial A_0(n)}, 
\eeq
where the first term represents the force coming from the gauge action, and the second term represents the force coming from the fermion action. Expressing the fermion action in terms of $\phi, \phi^{\dagger}$ one has
\beq
\label{FermionForce}
F^{(f)}_n = - \frac{\partial}{\partial A_0(n)} \left( \phi^{\dagger} \left( M^{\dagger} M \right)^{-1}  \phi \right), 
\eeq
where the calculation of the derivative with respect to $A_0(n)$ involves a significant number of terms for the asqtad action. The calculation of the fermion force also involves solving the sparse linear system $\left( M^{\dagger}M \right) X = \phi$, which is performed with an iterative solver.
The gauge force has a concise expression which can be obtained by differentiating (\ref{NCGaugeAction1}) with respect to $\theta_n$, which leads to
\beq
\label{GaugeForce}
F^{(g)}_n = -\beta \sum_i \left( 2A_0(n) - A_0(n+\hat{i}) - A_0(n-\hat{i}) \right).
\eeq
After each trajectory, the pseudofermion field $\phi$ and the gauge momenta are refreshed. This strategy has been known to improve the amount of phase space explored by the algorithm and goes
under the name of refreshed HMC.

In our simulations we have taken the trajectory length to be $1$, and varied the step size for a given ensemble in order to obtain a Metropolis acceptance rate of $70 \%$. We have discarded the first $200-250$ trajectories in order to account for the equilibration of each ensemble. For ensembles where we were interested in the mass spectrum of the theory,
we have generated roughly $800$ independent configurations. For those where we were interested only in the condensates, we generated roughly $100$ independent configurations. To achieve statistical independence, we have saved configurations after every tenth trajectory. An analysis of the autocorrelation function for various observables confirms that these configurations are sufficiently statistically independent.
\section{\label{sec:Observables}Observables}
The primary observable of this study is the chiral condensate. In the massless limit, this observable serves as an order parameter for the transition between the insulator phase, where it is nonzero, and the semimetal phase, where it vanishes.
In the continuum graphene EFT, the chiral condensate is defined as
\beq
\label{ChiralCondensateContinuum} \nn
\vev{\Bpsi \Psi} &=& \frac{1}{V}\frac{\partial \log Z}{\partial m}  \\ &=&  \frac{1}{V} \frac{1}{Z} \int \mathcal{D}A_0  \tr \left(\Dslash + m \right)^{-1}e^{-S^{\text{eff}}_E[A_0]},
\eeq
where $S^{\text{eff}}_E\left[A_0\right] = S_G\left[A_0\right] - \tr \log\left(\Dslash + m\right)$, and the partition function is given by
\beq
\label{PartitionFunction}
Z = \int \mathcal{D}A_0 \mathcal{D} \Bpsi \mathcal{D} \Psi e^{-S_E\left[A_0, \Bpsi, \Psi \right]}.
\eeq
The fermions have been integrated out of the partition function using the identity
\beq
\det \left( \Dslash + m \right) = \int \mathcal{D} \Bpsi \mathcal{D} \Psi e^{-\Bpsi \left( \Dslashexp + m \right) \Psi}.
\eeq
For staggered fermions in the one-component basis, the scalar density takes the form $\chib \chi$, and thus the chiral condensate becomes
\beq
\label{ChiralCondensateLattice}
\vev{ \chib \chi } = \frac{1}{V} \frac{1}{Z} \int \mathcal{D}A_0  \tr \left(\Dslash_{st} + m \right)^{-1}e^{-S^{\text{eff}}_E[U_0]},
\eeq
where in our lattice partition function we integrate over the noncompact gauge potential $A_0$.
In order to determine if magnetic catalysis is present in the graphene EFT, we will be interested in taking the zero-temperature as well as massless limit for the chiral condensate.

Another condensate which characterizes magnetic catalysis in the graphene EFT is the Haldane condensate. This condensate is time-reversal-odd and takes the form
\begin{widetext}[
\beq
\label{HaldaneCondensateContinuum}
\Delta_H \equiv \vev{\Bpsi \left( \tilde{\gamma}_{4,5} \otimes \bm 1 \right) \Psi}  = \frac{1}{V} \frac{1}{Z} \int \mathcal{D}A_0  \int d^3x \Bpsi(x) \left( \tilde{\gamma}_{4,5} \otimes \bm 1 \right) \Psi(x)  e^{-S^{\text{eff}}_E[A_0]}.
\eeq
]\end{widetext}
In the one-component basis, this takes the form
\beq
\label{HaldaneCondensateLattice}
\Delta_H = \frac{1}{V} \frac{1}{Z} \int \mathcal{D}A_0 \sum_{y;\eta_{\mu} \neq \eta'_{\mu}} \chib_{\eta}(y) \chi_{\eta'}(y)e^{-S^{\text{eff}}_E[U_0]},
\eeq
where the details of the transformation connecting (\ref{HaldaneCondensateContinuum}) and (\ref{HaldaneCondensateLattice}) are given in Appendix \ref{sec:SpinTasteAppendix}.
As opposed to (\ref{ChiralCondensateLattice}), the fields in (\ref{HaldaneCondensateLattice}) are located at diagonally opposite corners of the $2^3$ cube.
Although the appearance of a nonzero value for this condensate does not signal spontaneous symmetry breaking of the $U(1)_{\epsilon}$ symmetry, its value and dependence on the external magnetic field characterize the ground state of the graphene EFT.

To measure condensates on the lattice, one must use stochastic methods even for modest lattice volumes. For example, the trace in (\ref{ChiralCondensateLattice}) involves computing the propagator from a given site back to the same lattice site for each site on the lattice. To estimate the trace, we generate an ensemble of Gaussian-distributed complex vectors with support on each lattice site that satisfy
\beq
\label{StochasticOrthogonality}
\vev{ \Phi^*_i \Phi_j } = \delta_{ij},
\eeq
where the expectation value is over the Gaussian distribution and $i$ and $j$ are lattice-site labels. We then can obtain
\beq
\tr M^{-1} \approx \frac{1}{N_v} \sum^{N_v}_{k=1} \Phi^{(k) \dagger} M^{-1} \Phi^{(k)},
\eeq
where $N_v$ is the number of stochastic vectors chosen from the distribution. Typically, to obtain a good estimate of the chiral condensate, we need approximately $100$ stochastic vectors per configuration. 

To estimate the condensate in (\ref{HaldaneCondensateLattice}), one must alter the procedure done for the chiral condensate as the fields reside at opposite corners of the cube. We first generate an ensemble of Gaussian-distributed complex vectors satisfying (\ref{StochasticOrthogonality}) with support only on sites with a given $\eta = (\eta_t, \eta_x, \eta_y)$, where $\eta_{\mu}$ labels a particular site within the cube. We then shift the source to the opposite corner of the cube using parallel transport
\beq
\label{ShiftedStochasticSource}
\tilde{\Phi}^{(k)} = \frac{1}{6} \sum_{P} \hat{S}_{P_{\mu}} \hat{S}_{P_{\nu}} \hat{S}_{P_{\lambda}} \Phi^{(k)},
\eeq
 where the sum is over the six permutations taking us to the corner opposite $\eta$. Here $P_{\mu}, P_{\nu}, P_{\lambda}$ are permutations of $\pm t, \pm x, \pm y$, and we have introduced the shift operator
 \beq
\label{ShiftOperator} 
 \left(\hat{S}_{\pm \mu} \Phi \right)_i &=& \left\{ \begin{array}{ll}  U^{\dagger}_{\mu}(x - \hat{\mu}) \Phi_{i-\hat{\mu}} &, - \\                    U_{\mu}(x) \Phi_{i+\hat{\mu}} &, +
                   \end{array} \right. .
 \eeq
 Using the source in (\ref{ShiftedStochasticSource}), we can calculate the Haldane condensate as follows
 \beq
 \label{HaldaneStochastic}
 \sum_{y;\eta_{\mu} \neq \eta'_{\mu}} \chib_{\eta}(y) \chi_{\eta'}(y) \approx \frac{1}{N_v}  \sum^{N_v}_{k=1} \Phi^{(k) \dagger} M^{-1} \tilde{\Phi}^{(k)}.
 \eeq
where the sum over $\eta $ goes over all eight sites of the cube. Stochastic estimation relies on cancellations of the noise to find the signal, which decays exponentially with the separation between source and sink.  Therefore, for operators such as the Haldane condenstate, which are nonlocal, it can be difficult to obtain a good signal-to-noise ratio. In order to overcome this, we have taken $N_v \approx 1000$ in (\ref{HaldaneStochastic}). 

 
The spontaneous breaking of the $U(1)_{\epsilon}$ symmetry also has observable consequences for the fermion quasiparticle, which receives a dynamical mass, $m_F$, which remains nonzero even as the bare mass vanishes.
In order to study the dynamical fermion mass, we calculate the quasiparticle propagator in the temporal direction
\beq
\label{QuasiparticlePropagatorTemporal}
G^{(\tau)}_F(\tau; \vec{p}) = \sum_{\vec{r}} e^{i \vec{p} \cdot \vec{r}}\vev{\chi(x,y,\tau) \chib(0,0,0)},
\eeq
where $\vec{p} \equiv (p_x,p_y)$ and $\vec{r} \equiv (x,y)$. In order to investigate effects of the finite spatial size of the box, we are also interested in the quasiparticle propagator in the spatial direction 
\beq
\label{QuasiparticlePropagatorSpatial}
G^{(s)}_F(x; \omega_l, p_y) = \sum_{\tau,y} e^{i \left( \omega_l \tau + p_y y \right)} \vev{\chi(x,y,\tau) \chib(0,0,0)},
\eeq
where $\omega_l = (2l+1)\pi T,~l=0,1,2,\dots$, are the fermionic Matsubara frequencies.

As a result of the spontaneous breaking of the $U(1)_{\epsilon}$ symmetry on the lattice, there should arise a pseudoscalar NG boson. This state is analagous to
the pion in QCD. As in QCD, one expects this state to become massless as the bare mass vanishes. Its mass, $m_{\pi}$, is studied through the following two-point correlator in Euclidean time
\beq
\label{PionTemporal}
G^{(\tau)}_{PS}(\tau; \vec{p}) = \sum_{x,y} e^{i \vec{p} \cdot \vec{x}} \vev{ \mathcal{O}_{PS}(x,y,\tau) \mathcal{O}_{PS}(0,0,0) },
\eeq
where $\mathcal{O}_{PS}$ is a staggered bilinear operator with pseudoscalar quantum numbers. In the spin-taste basis, the operator we have used is
\beq
\label{PSSpinTaste}
\mathcal{O}_{PS}(y) = \Bpsi(y) \left(\tilde{\gamma}_4 \otimes \bm 1 \right)\Psi(y).
\eeq
In the one-component basis, this operator becomes
\beq
\label{PS1Comp}
\mathcal{O}_{PS}(y) = \sum_{\eta} \epsilon(x) \chib_{\eta}(y) \chi_{\eta}(y),
\eeq
where $x_{\mu} = 2y_{\mu} + \eta_{\mu}$, $\epsilon(x) = (-1)^{\eta_0 + \eta_1 + \eta_2}$, and the sum runs over the sites of the cube labeled by $y$. The details of the transformations connecting (\ref{PSSpinTaste}) and (\ref{PS1Comp}) are given in Appendix \ref{sec:SpinTasteAppendix}.
The spatial pseudoscalar correlator is also of interest and is given by 
\beq
\label{PionSpatial}
\nn
G^{(s)}_{PS}(x; \omega_l, p_y) &=& \sum_{x,y} e^{i \left( \omega_l \tau + p_y y \right)} \\  &\times& \vev{ \mathcal{O}_{PS}(x,y,\tau) \mathcal{O}_{PS}(0,0,0) },
\eeq
where $\omega_l = 2l\pi T,~l=0,1,2,\dots$, are the bosonic Matsubara frequencies.
\section{\label{sec:Results}Results}
\subsection{\label{sec:ChiralCond}Chiral Condensate}
In this section we will discuss the results of our calculations regarding magnetic catalysis in the graphene EFT. Using the lattice methods described above, we will attempt to characterize the symmetry breaking using various observables.
Apart from the condensates defined above, we will also investigate the quasiparticle and pseudoscalar mass. Thus, various methods which have been used to study chiral symmetry breaking in the
context of QCD can be applied to the graphene EFT.

In the absence of an external magnetic field, the graphene EFT is known to exhibit two phases: a semimetal phase at weak coupling (large dielectric constant $\epsilon$) and an insulating phase at strong coupling (small $\epsilon$). 
In the language of chiral-symmetry breaking, the semimetal phase corresponds to the chirally symmetric phase while the insulating phase corresponds to the broken phase.
Various aspects of this transition were studied previously using lattice methods \cite{Drut1, Drut2, Hands1, Giedt}. The transition is believed to be of second order as indicated by the results in \cite{Drut2}. 

The situation changes as one turns on the external magnetic field. Namely, the critical $\beta_c = (\epsilon_c+1)/2e^2$ which determines the boundary between the two phases shifts to larger values. This has been shown previously in \cite{Polikarpov}, where at fixed temperature, the authors
obtained a phase diagram in the $(B, \beta)$-plane. One would expect that the phase boundary has a temperature dependence, where at $T=0$, the authors of \cite{Miransky1,Miransky2,Miransky3,Miransky4,MiranskyGraphene1,MiranskyGraphene2,MiranskyGraphene3} predict that an infinitesimal attraction between fermions and antifermions will lead to pairing, and thus magnetic catalysis. Another early analysis of a graphene-like theory showed that at extremely weak coupling, a nonzero condensate was obtained in the chiral limit \cite{Cosmai}

\begin{figure}
\vspace{-1.2cm}
\includegraphics[height=9cm,width=9cm]{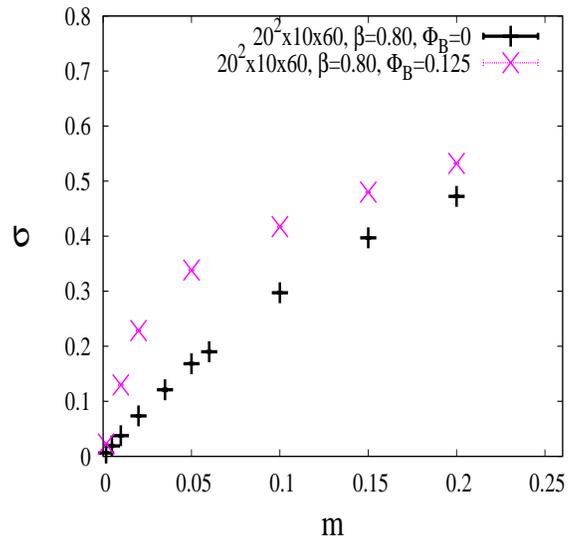} 
\vspace{-1.25cm}
\caption{$\sigma \equiv \vev{\Bpsi \Psi}$, the chiral condensate (in units of $a^2_s$), as a function of the bare fermion mass (in units of $a_t$) at zero external magnetic field (black points) and at magnetic flux $\Phi_B=0.125$ (pink points). Here, the magnetic flux is measured in units of $a^2_s$. The volumes are reported in the form $N^2_s \times N_z \times N_{\tau}$.  We note that $\sigma$ vanishes with $m$ at nonzero field as well as at zero external magnetic field. Thermal effects are argued to be the reason behind the vanishing of the condensate in the presence of the magnetic field. The statistical errors on each point are not visible on this scale.}
\label{PBPComparison}
\end{figure}

We began our calculations by identifying the semimetal phase through a scan in $\beta$, coupled with an investigation of the condensate's behavior as a function of the bare mass in the chiral limit. In the semimetal phase, one expects $\sigma \equiv \vev{\Bpsi \Psi}$ to be linear in the bare mass for small values of $m$. 
One can compare this expectation with the behavior of $\vev{\Bpsi \Psi}$ as a function of the bare mass when the external magnetic field is turned on. This is illustrated in Fig.~\ref{PBPComparison}, where one sees an approximately linear behavior at zero field and a strongly nonlinear behavior of the condensate as a function of $m$ for the magnetic flux $\Phi_B = 0.125$.

\begin{figure}
\vspace{-1.2cm}
\includegraphics[height=9cm,width=9cm]{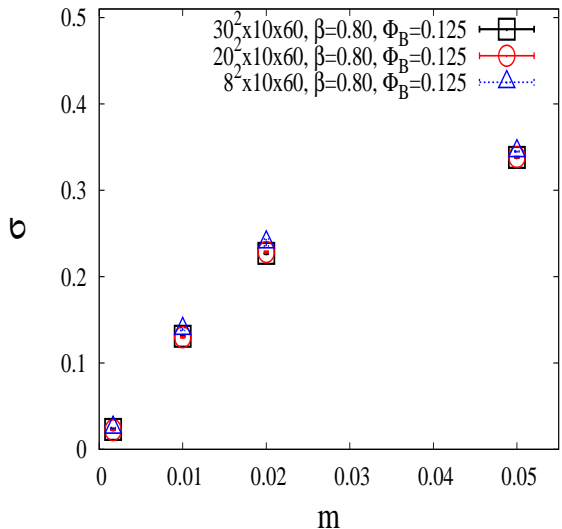} 
\vspace{-1.25cm}
\caption{The chiral condensate as a function of the fermion mass for different spatial volumes $N^2_s$ at magnetic flux $\Phi_B=0.125$. The volumes are listed in the form $N_s^2\times N_z \times N_{\tau}$, where the fermions live in the $xy$-plane and the gauge field is present throughout the entire volume.}
\label{PBPVolume}
\end{figure}

Notice, however, that in both cases, as the explicit symmetry breaking parameter is taken to zero, the condensate also goes to zero. Even if SSB occurs, this still may occur due to the finite spatial volume. Referring to Fig.~\ref{PBPVolume}, one can see that this is not the
case. For a broad range of spatial extents $N_s$, the condensate shows little variation. This can be explained by observing that the magnetic length, $l_B \equiv \sqrt{\hbar c/eB}$, which characterizes the fermion's cyclotron orbit, satisfies $1 < l_B < N_s$, in units of $a_s$. 

\begin{figure}
\vspace{-1.2cm}
 \includegraphics[height=9cm,width=9cm]{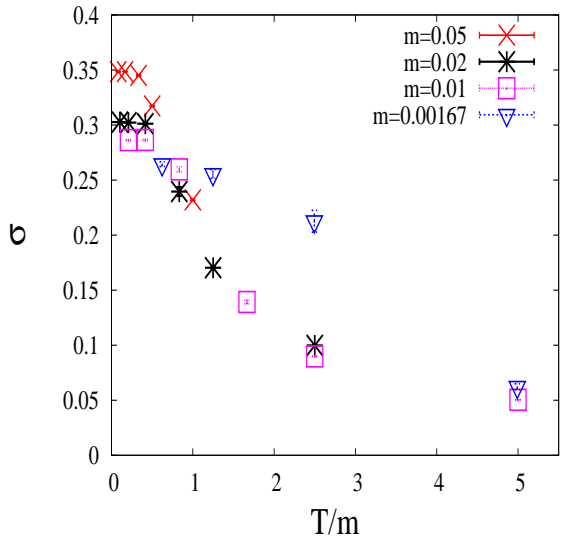} 
 \vspace{-1.25cm}
\caption{The chiral condensate plotted as a function of the ratio $T/m$ for the ensembles with $\Phi_B=0.125$ and $N_s=8$. One sees that at small values of $T/m$, the condensate increases and plateaus towards a nonzero value.}
\label{PBPvsTdivM}
\end{figure}

\begin{figure}
\vspace{-1.2cm}
  \includegraphics[height=9cm,width=9cm]{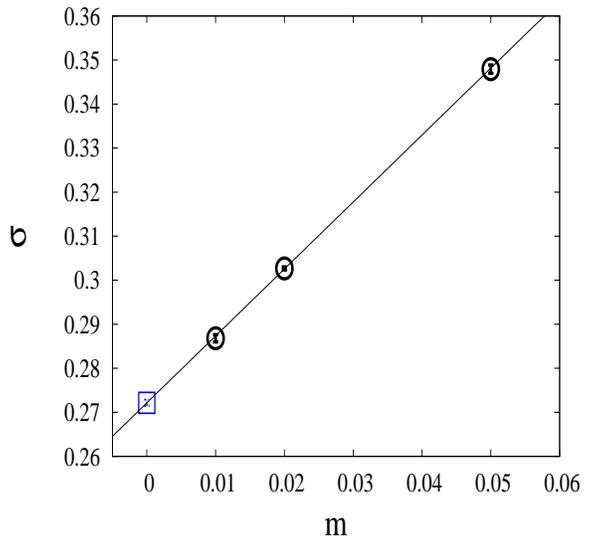} 
  \vspace{-1.25cm}
\caption{The chiral limit of the chiral condensate using the points extrapolated to $T=0$. The error bars on each point were determined from both systematics i.e. choice of model for the zero-temperature extrapolation, as well as statistics. The value of the intercept for the linear fit is $0.2721(7)$ ($\chi^2/\text{d.o.f.} \approx 0.6$).}
\label{PBPzeroTChiral}
\end{figure}

\begin{figure}
\vspace{-1.2cm}
 \includegraphics[height=9cm,width=9cm]{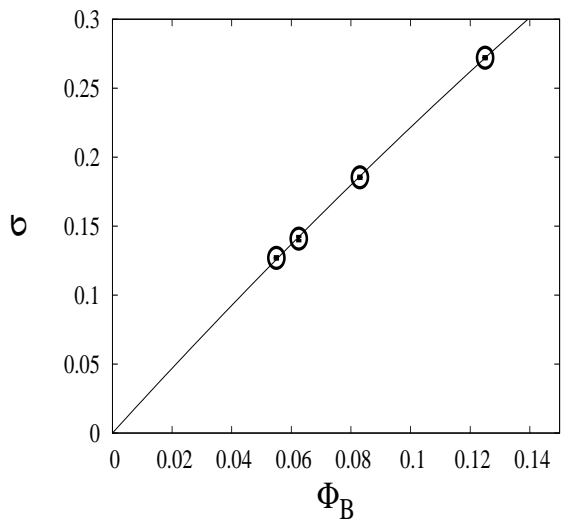} 
 \vspace{-1.25cm}
\caption{The zero-temperature, chirally-extrapolated values of the condensate, plotted as a function of the external magnetic flux,  $\Phi_B = eB/2\pi$.  The points at $\Phi_B=0.083$ and $\Phi_B=0.056$ have a spatial size of $N_s=12$, while those at $\Phi_B=0.125$ and $\Phi_B=0.0625$ have a spatial size of $N_s=8$. The error bars on the points come from the chiral extrapolations at $T=0$. The data have been fit to a quadratic constrained to pass through the origin ($\chi^2/\text{d.o.f.} \approx 3.6/2$). } 
\label{PBPzeroTChiralvseB}
\end{figure}

We have also confirmed the independence of the results on the finite spatial size of the box by indepedently calculating the screening
masses for the fermion quasiparticle and the pseudoscalar. The screening masses are obtained by calculating the correlation functions in (\ref{QuasiparticlePropagatorSpatial}) and (\ref{PionSpatial}) and projecting to $p_y=0$ and the lowest Matsubara frequency ($\omega_0 = \pi T$ and $\omega_0 = 0$, respectively).
We find that the masses characterizing the decay of the above correlation functions in space satisfy $M_s L_s \gg 1$, where $L_s = N_s a_s$ is the  lattice extent in the spatial direction. For the pseudoscalar, we found that $m_{\pi,s}L_s \approx 18-20$, for the ensembles with volume $20^2\times10\times60$ and flux $\Phi_B=0.125$. For the fermion propagator, we obtained $m_{F,s}L_s \approx 11-14$, for the same ensembles mentioned above. This is further proof that the corrections arising from the finite spatial extent of the box are well under control. The dependence of the condensate on $N_z$, the spatial direction perpendicular to the plane of graphene, was also checked. We found that for $N_z \geq 10$, the correction to the infinite-volume result is less than $2\%$.
 
Thermal effects are also known to play a role in symmetry restoration. The temperature of the system is related to the extent of the lattice in Euclidean time, $T = 1/N_{\tau}a_t$. We have investigated the effect of finite temperature on the condensate and illustrated it in Fig.~\ref{PBPvsTdivM}, where $\sigma(T/m)$ is plotted. When the temperature is large 
compared with the fermion mass, one can see that the condensate tends towards zero. On the other hand, when the temperature is small compared with the fermion mass, one can see that the condensate increases and tends asymptotically towards a nonzero value. By first taking the limit $T \to 0$, we are able to study magnetic catalysis in the ground state. Using this strategy, we have first extrapolated to zero temperature at fixed bare fermion mass. We have extrapolated using two different methods. First, we have taken points at small $T$ which form a plateau and fit them to a constant. These plateaus can be seen in Fig. \ref{PBPvsTdivM}, for magnetic flux $\Phi_B=0.125$. We have then included one other point at larger $T$ and extrapolated using a quadratic fit (except for bare mass $m=0.01$ in Fig. \ref{PBPvsTdivM}, where the third point is too far away from $T=0$). We use the difference between the two results for $\sigma$ at $T=0$ to estimate a systematic uncertainty associated with this extrapolation, with the central value taken as the average. A similar procedure has been also carried out for the other three magnetic fluxes. We then use these points for our chiral extrapolation, which we performed using a linear fit. This is illustrated in Fig. \ref{PBPzeroTChiral}, where we plot the mass dependence of the $T=0$ extrapolated points.

Proceeding in an analogous manner for three other fluxes, we were able to obtain the behavior of the zero-temperature, chirally extrapolated condensate as a function of the magnetic flux, $\Phi_B$. These results are shown in Fig.~\ref{PBPzeroTChiralvseB}.
The relationship between the condensate and the magnetic flux is fit to the form $\sigma_{T=m=0} = eBc_1 + (eB)^2c_2$. The errors on the points are those calculated in the chiral extrapolation. These results overwhelmingly support the scenario of magnetic catalysis in the graphene EFT.

\subsection{\label{sec:HaldaneCond}Haldane Condensate}
Long before the realization of graphene in the lab, F.D.M. Haldane considered a graphene-like lattice model which he used to study the quantum Hall effect (QHE) \cite{Haldane}.
In this model, he considered a magnetic field which was periodic in space such that the flux through the unit cell was zero. This is in stark contrast to normal realizations of the QHE 
where the electrons are subjected to a uniform magnetic field, which implies the formation of Landau levels. In Haldane's model, the electrons retain their Bloch character, and thus it is referred to 
as the QHE without Landau levels. As in graphene, the points in his model where the valence and conductance bands touch inside the Brillouin zone are symmetry-protected (inversion and time-reversal).
If time-reversal invariance is broken, he found that a $\nu = \pm 1$ integer QHE state is formed, where $\nu$ is referred to as the filling factor. 

Recent studies of symmetry breaking in graphene have brought up the possbility of the formation of the time-reversal-odd condensate, $\Delta_H$ \cite{GonzalezHaldaneMass, MiranskyGraphene2,MiranskyGraphene3}. This condensate gives rise to the so-called Haldane mass in the low-energy theory.
The reason for the interest in these symmetry breaking scenarios is as follows.
The appearance of anomalous QHE states in graphene at previously unknown filling factors, $\nu = 0, \pm 1, \pm 2$, is due to the splitting of the degeneracy of the lowest Landau level (LLL) at large external magnetic fields ($B \sim 45 \text{T}$).
Although there are several scenarios to explain these states, (see \cite{Yang} for a nice discussion), the authors of \cite{MiranskyGraphene2, MiranskyGraphene3} advocate the magnetic catalysis scenario whereby a Dirac mass as well as possibly a Haldane mass are generated due to the strong, weakly-screened
electron-electron interactions in graphene.

In the Schwinger-Dyson approach employed in \cite{MiranskyGraphene2}, where a simplified contact interaction was used, the ground state of the EFT in the presence of an external magnetic field was found to be described by
\beq
\label{Singlet}
\tilde{\Delta}_{\uparrow} = \tilde{\Delta}_{\downarrow} = 0, ~\Delta_{H,\uparrow} = - \Delta_{H,\downarrow} = M,
\eeq
where $\tilde{\Delta}_{\sigma}$ corresponds to the Dirac mass for a given spin projection, and $\Delta_{H,\sigma}$ corresponds to the Haldane mass for a given spin projection. In this expression, $M$ is the dynamically generated mass scale which is a function of the cutoff $\Lambda$ and the dimensionful coupling constant characterizing the contact interaction, $G$. We note that the spin index in (\ref{Singlet}) directly corresponds to the taste index for staggered fermions. In \cite{MiranskyGraphene3}, a more realistic long-range Coulomb interaction in the instantaneous approximation was used. The results were similar to that of \cite{MiranskyGraphene2}, with the qualitative difference that the gap parameters depend on the Landau-level index.

Although we ignore it in our lattice simulations, the Zeeman term plays a significant role in the selection of the ground state.  According to \cite{MiranskyGraphene2,MiranskyGraphene3}, the solution 
\beq
\label{Triplet}
\tilde{\Delta}_{\uparrow} = \tilde{\Delta}_{\downarrow} = M, ~\Delta_{H,\uparrow} = \Delta_{H,\downarrow}=0,
\eeq
is degenerate with (\ref{Singlet}) in the absence of Zeeman splitting. This makes the comparison with continuum results more difficult and is an avenue for future work.

In our simulations, we have used operators which are all taste singlets and thus have not distinguished the various ``spin" components. Taste-nonsinglet operators have zero expectation value as the vacuum does not have a preferred direction in taste space unless one introduces a taste-breaking term into the action. Thus, one is unable to project out the various spin components of the order parameters, which makes a comparison with the continuum results much more difficult.  In light of this fact, we have looked at the taste-singlet Haldane mass characterized by the operator $\Bpsi \left(\tilde{\gamma}_{4,5} \otimes \bm 1\right) \Psi$. In contrast with the chiral condensate, we have not added a term like this to the action and thus cannot perform the same analysis of removing the explicit mass term after taking the infinite volume limit. Using lattices generated with the action (\ref{GrapheneFermLatticeDimensionless}), we can look at the ensemble average of the operator as well as look for the appearance of a first-order phase transition in the time history of the taste-singlet operator characterized by a tunneling between negative and positive values.

\begin{figure}
\vspace{-1.2cm}
 \includegraphics[height=9cm,width=9cm]{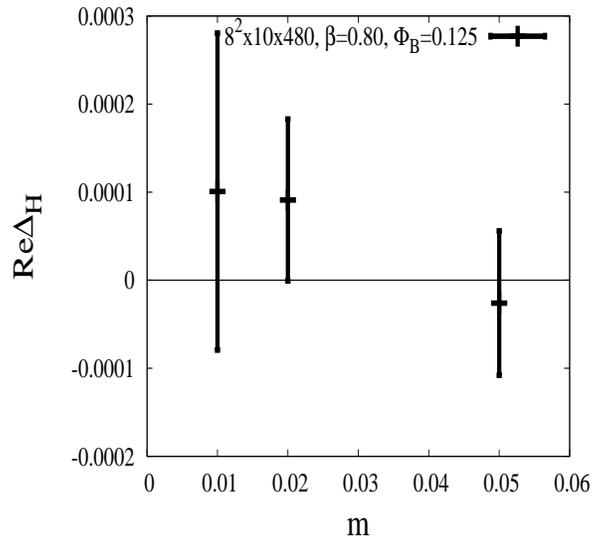} 
 \vspace{-1.25cm}
\caption{The real part of the Haldane mass as a function of the fermion bare mass $m$ for $8^2\times10\times480,\beta=0.80, \Phi_B=0.125$.}
\label{ReHaldanevsm} 
\end{figure}

\begin{figure}
\vspace{-1.2cm}
 \includegraphics[height=9cm,width=9cm]{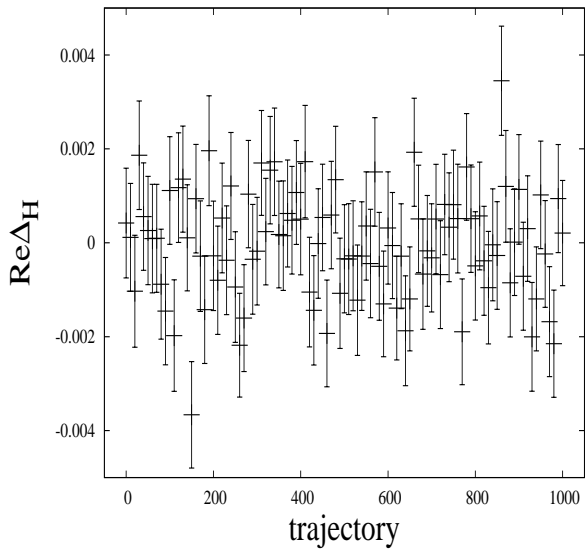} 
 \vspace{-1.25cm}
\caption{Monte Carlo time history of the real part of the Haldane mass  for $8^2\times10\times480,\beta=0.80, m=0.01, \Phi_B=0.125$. The errors on the points are the standard deviation of the mean of the stochastic estimation of the Haldane mass. }
\label{ReHaldaneTS} 
\end{figure}

We have measured the taste-singlet time-reversal-odd condensate on the lattice ensembles with the largest flux $\Phi_B=0.125$ for $a_t T = [0.002, 0.016]$. These values of $a_tT$ are smaller than the dynamical fermion mass, as one can see from Fig. \ref{MFvsmNonzeroB}. Thus, one expects that if the ground state did support a Haldane condensate, these temperatures would be sufficiently small and the magnetic flux sufficiently large to observe this. Instead, we find that our data do not support a nonzero taste-singlet Haldane condensate. In Fig.~\ref{ReHaldanevsm}, we have plotted the real part of the Haldane mass, computed with $1000$ stochastic sources on $100$ gauge configurations, versus $m$ for $N_{\tau}=480,~\Phi_B=0.125$. Our results are consistent with zero. We have also checked, at the same value of the flux and bare mass, that as the temperature increases, the Haldane mass remains consistent with zero as to be expected from the solutions of the gap equations. In Fig.~\ref{ReHaldaneTS}, we plot the Monte Carlo time history of the taste-singlet Haldane mass. The values are regularly distributed around the mean and do not show any indication of tunneling, which would characterize a first-order phase transition. Thus we do not see any evidence for the formation of a taste-singlet Haldane condensate in our calculations. Keeping in mind the limitations of our approach, one would need to perform further simulations with an explicit taste-singlet mass in the action in order to test for spontaneous breaking of the discrete symmetry.

\subsection{\label{sec:DiracQuasi}Dirac Quasiparticle}
We also studied the fermion quasiparticle propagator. Due to the appearance of a nonzero value for the condensate $\vev{\Bpsi \Psi}$, one expects that the dynamical mass remains nonzero in the limit that the bare mass vanishes. 
Although the propagator itself is gauge variant, the pole of the propagator in Coulomb gauge is a familiar quantity. In order to extract information from the quasiparticle propagator, we apply a transformation on the temporal links. Our gauge fixing procedure sets the average value of the potential on a given time slice to zero, 
\beq
\sum_{x,y} A_0(x,y,\tau) = 0.
\eeq
This allows us to calculate the propagator of charged particles, as the gauge condition corresponds to a charged propagator moving in the presence of a spatially uniform background charge of opposite sign.
To determine the dynamical fermion mass, $m_F$, we fit the propagator in (\ref{QuasiparticlePropagatorTemporal}) to the following form
\beq
\label{FermFitForm} 
G_F(\tau, \vec{p} = {\bf 0}) = A\left( e^{-m_F\tau} + (-)^{\tau}e^{-m_F(N_{\tau} - \tau)} \right).
\eeq

The fits that we performed for the quasiparticle propagator necessitated proper care in order to accurately determine the ground state mass. To do this one must select a fit range, $[ \tau_{\text{min}}, \tau_{\text{max}}]$, where the correlator is free of excited-state contributions. We have found that the selection of $\tau_{\text{min}}$ primarily depends on the value of the external magnetic flux, $\Phi_B$. This is due to the fact that the excited state with the same oscillating behavior as the ground state consists of the fermion promoted to the next highest LL. One can also have an excited state consisting of a fermion along with a pseudoscalar particle having nonzero orbital angular momentum. Our result for particle masses suggest that this contribution dies out quite quickly in Euclidean time, $\tau$.

\begin{figure}
\vspace{-1.2cm}
  \includegraphics[height=9cm,width=9cm]{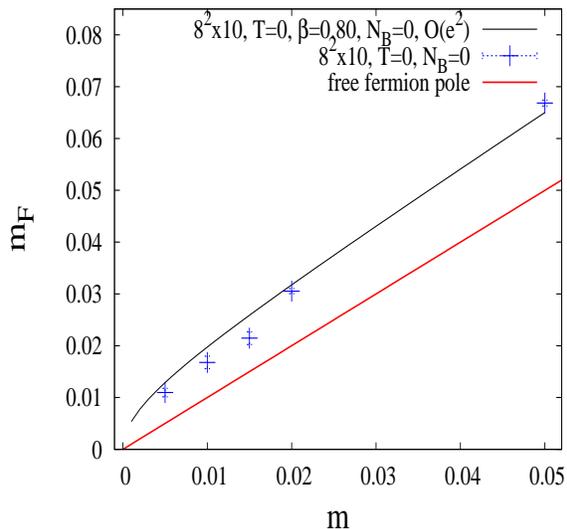} 
  \vspace{-1.25cm}
\caption{The fermion dynamical mass as a function of the bare fermion mass for zero magnetic flux. We have included the result for the pole of the fermion propagator at $O(e^2)$ (upper curve) along with the free fermion pole, $\log\left(m+\sqrt{m^2+1}\right)$ (lower curve). One can see that the perturbative result is close to our nonperturbative result.}
\label{MFvsmZeroB}
\end{figure}

\begin{figure}
\vspace{-1.2cm}
  \includegraphics[height=9cm,width=9cm]{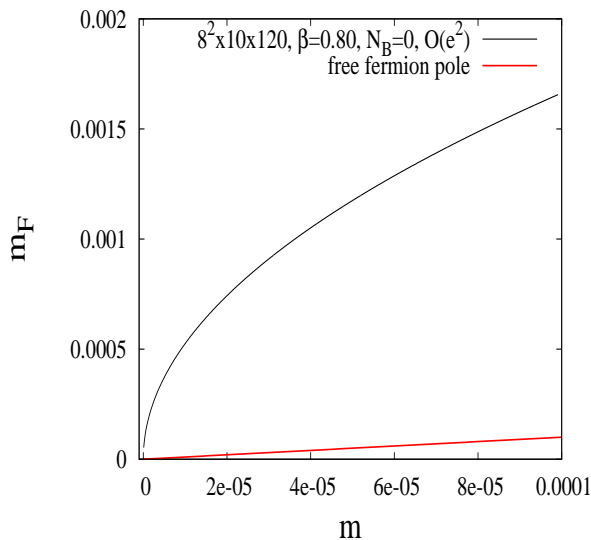} 
  \vspace{-1.25cm}
\caption{The pole of the fermion propagator at $O(e^2)$ (upper curve) along with the free fermion pole (lower curve) as one approaches the chiral limit, $m \to 0$. One can see that the pole at $O(e^2)$ vanishes in this limit, as expected. However, the curvature that causes this behavior can be observed only as one moves to extremely small fermion bare masses.}
\label{MFvsmZeroBZoom}
\end{figure}

At zero magnetic field, we expect the dynamical fermion mass to vanish in the chiral limit. In Fig.~\ref{MFvsmZeroB}, one can see our nonperturbative determination of the dynamical fermion mass. However, it appears that making an extrapolation to zero bare mass, based on the nonperturbative results, leaves us with a nonzero value for the dynamical fermion mass. According to the arguments and results previously discussed, one expects the dynamical mass to vanish in this limit as we are firmly in the semimetal phase in the absence of the external magnetic field. To investigate this result further, we calculated the pole position of the lattice fermion propagator at $O(e^2)$ using lattice perturbation theory. In this calculation we employed a one-link staggered action with tadpole improvement determined at $O(e^2)$. As one can see in Fig.~\ref{MFvsmZeroB}, the perturbative result exhibits a significant renormalization of the mass compared with the free theory. Furthermore, referring to Fig.~\ref{MFvsmZeroBZoom}, one can see that as one approaches the origin, the perturbative result shows large curvature and eventually vanishes at the origin. Using this result as a heuristic explanation of our results at zero magnetic field, one might expect the same to happen in the chiral limit for a nonperturbative calculation.

\begin{figure}
\vspace{-1.2cm}
  \includegraphics[height=9cm,width=9cm]{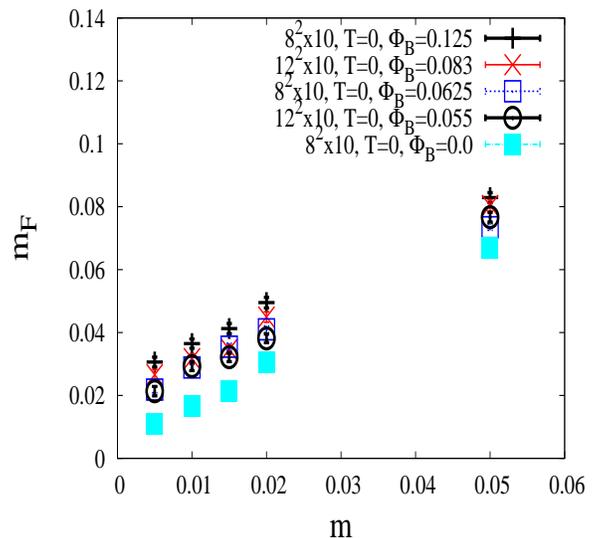} 
  \vspace{-1.25cm}
\caption{The dynamical fermion mass plotted as a function of the bare mass for zero external magnetic flux and all four nonzero external magnetic fluxes. We have chosen $N_{\tau}=120,240,240,240,480$ for the bare fermion masses $m=0.05,0.02,0.015,0.01,0.005$, respectively.}
\label{MFvsmNonzeroB}
\end{figure}

Fig.~\ref{MFvsmNonzeroB} shows results for the dynamical fermion mass with nonzero and zero magnetic flux. One sees that at a given bare mass, the dynamical mass increases with the magnetic flux. Furthermore, the plot suggests that the values for all four nonzero magnetic fluxes extrapolate to nonzero values in the chiral limit. However, in light of the behavior observed in the zero field case, one might want to be cautious in predicting the behavior at bare masses smaller than those plotted. For all of the ensembles depicted in Fig.~\ref{MFvsmNonzeroB}, we have chosen an $N_{\tau}$ which corresponds to a temperature where the chiral condensate plateaus. Thus, we expect that we are effectively at $T=0$ when determining the dynamical mass.

\begin{figure}
\vspace{-1.2cm}
  \includegraphics[height=8.5cm,width=8.5cm]{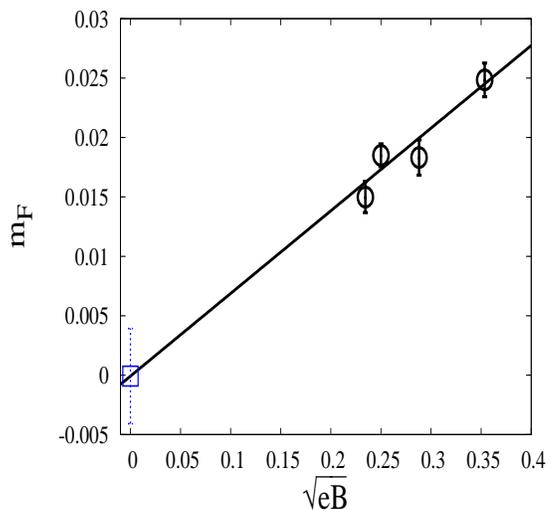} 
  \vspace{-1.25cm}
\caption{The $T=0$, chirally-extrapolated dynamical fermion mass as a function $\sqrt{eB}$. We have depicted a linear fit which gives an intercept which is consistent with zero ($\chi^2/\text{d.o.f.} \approx 3.7/2$). This confirms previous perturbative predictions.}
\label{fig:MFvsSqrtB}
\end{figure}

As previously mentioned, perturbative approaches to magnetic catalysis predict that for $(2+1)$-dimensional field theories, the dynamical fermion mass scales linearly with $\sqrt{eB}$ \cite{Shovkovy}. In order to check this, one first needs to extrapolate the dynamical mass to the chiral limit. Our results, depicted in Fig.~\ref{fig:MFvsSqrtB}, show that the chirally extrapolated dynamical mass behaves as expected. This provides further evidence in favor of magnetic catalysis in the graphene EFT.

\subsection{\label{sec:PS}Pseudoscalar}
\begin{figure}
\vspace{-1.2cm}
  \includegraphics[height=9cm,width=9cm]{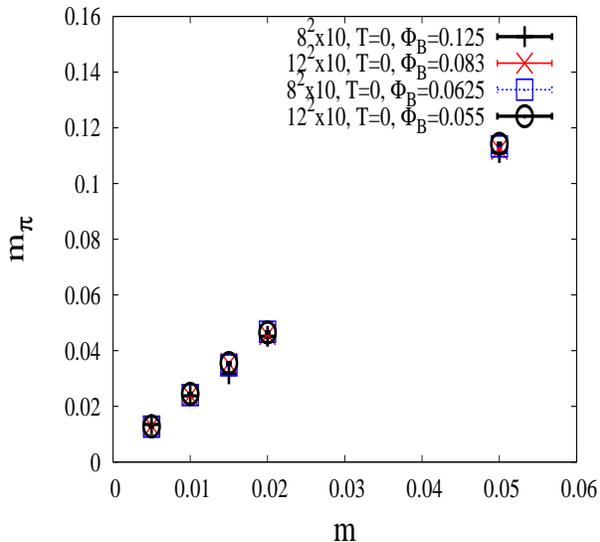} 
  \vspace{-1.25cm}
\caption{The mass of the pseudoscalar bound state as a function of the bare fermion mass for all four magnetic fluxes.}
\label{MPSvsm}
\end{figure}

The appearance of a pseudoscalar Nambu-Goldstone boson is expected as a consequence of the spontaneous chiral symmetry breaking. To determine this state's mass we fit the correlator in (\ref{PionTemporal}) to the following form
\begin{widetext}[
\beq
\label{PS2Point} 
G^{(\tau)}_{PS}(\tau; \vec{p} = {\bf 0}) = A\left( e^{-m_{\pi}\tau} + e^{-m_{\pi}(N_{\tau} - \tau)} \right) + A'\left( e^{-m'_{\pi}\tau} + e^{-m'_{\pi}(N_{\tau} - \tau)} \right),
\eeq
]\end{widetext}
where we have included two sets of exponentials characterizing the ground state and the first excited state respectively.
In Fig.~\ref{MPSvsm}, we see the ground state mass in the pseudoscalar channel, $m_{\pi}$, plotted as a function of the fermion mass for the various magnetic fluxes.  
We notice that the pseudoscalar mass shows little variation with magnetic flux, which is not surprising since the pseudoscalar carries no charge. This is in contrast with QCD, where electrically charged  pions couple to the 
external magnetic field \cite{SmilgaShushpanov}. We also have determined that the pseudoscalar mode is indeed a bound state. This can be seen by referring to Fig.~\ref{MFvsmNonzeroB} where the dynamical fermion mass is plotted as a function of the bare mass. The fermion-antifermion scattering state has energy $2m_F$, which is higher than $m_{\pi}$ for all simulated bare masses.



\section{\label{sec:Conclusion}Conclusion}

Through a thorough, fully nonperturbative study of the graphene EFT, we have shown the existence of spontaneous symmetry breaking due to an external magnetic field.
We have characterized the ground state of the system by performing a zero-temperature extrapolation of our observables. Furthermore, we have commented on the difficulties in studying 
the Haldane mass with staggered fermions. Although we found no evidence for the taste-singlet Haldane condensate, one must take into account the limitations of our lattice setup in this regard. The evidence for a dynamically generated Dirac mass for the quasiparticle, the NG boson, and the nonzero value for the chirally extrapolated, $T=0$ chiral condensate all show
that indeed, magnetic catalysis is occurring in the graphene EFT. Further studies of magnetic catalysis in the graphene EFT could address other questions, such as how the phase diagram presented in \cite{Polikarpov} changes as one varies the temperature.
This would necessarily involve a scan in $\beta$ which differs from our current study which was performed at fixed $\beta=0.80$. Another direction one could pursue would be to 
study the valley-sublattice and spin densities which are order parameters for quantum Hall ferromagnetism (QHF). In a study of the Schwinger-Dyson gap equations, it was found that 
the appearance of the condensates associated with magnetic catalysis and those associated with QHF share the same dynamical origin \cite{MiranskyGraphene2}.

\acknowledgements
This work was in part based on a variant of the MILC collaboration's public lattice gauge theory code. See {\bf http://physics.utah.edu/$\sim$detar/milc.html}.
Calculations were performed at the Center for High Performance Computing at the University of Utah, Fermi National Accelerator Laboratory, and the LOEWE-CSC high performance supercomputer of Johann Wolfgang Goethe-University Frankfurt am Main. We would like to thank HPC-Hessen, funded by the State Ministry of Higher Education, Research and the Arts, for programming advice. Numerical computations have also used resources of CINES and GENCI-IDRIS as well as resources at the IN2P3 computing facility in Lyon.
The authors would like to acknowledge discussions with Maksim Ulybyshev, Dima Pesin, and Eugene Mishchenko. SZ would like to acknowledge discussions with Wolfgang Unger.
SZ would like to acknowledge the support of the Alexander von Humboldt foundation. CW and CD were supported by the US NSF grant PHY10-034278.

\appendix
\section{\label{sec:SpinTasteAppendix}Spin-Taste Basis in $(2+1)$ dimensions}
In this appendix we discuss the spin-taste basis in $(2+1)$, dimensions which has some differences with the more familiar $(3+1)$ dimensional case. 
The discussion follows that of \cite{Burkitt}. Starting from the one-component fields $\chi, \chib$ one performs a change of basis by introducing the following transformation
\beq
\label{SpinTasteTransfo2+1a}
u^{\alpha a}(y) &=& \frac{1}{4\sqrt{2}} \sum_{\eta} \Gamma^{\alpha a}_{\eta} \chi_{\eta}(y), \\ 
\label{SpinTasteTransfo2+1b}
d^{\alpha a}(y) &=& \frac{1}{4\sqrt{2}} \sum_{\eta} B^{\alpha a}_{\eta} \chi_{\eta}(y),
\eeq
where one has introduced the matrices 
\beq
\label{SpinTasteMatrices2+1}
\Gamma_{\eta} &\equiv& \sigma^{\eta_0}_0 \sigma^{\eta_1}_1 \sigma^{\eta_2}_2, \\ ~B_{\eta} &\equiv& \beta^{\eta_0}_0 \beta^{\eta_1}_1 \beta^{\eta_2}_2, ~\beta_{\mu}\equiv - \sigma_{\mu}.
\eeq
We note that one labels a lattice site by $n_{\mu} = 2y_{\mu} + \eta_{\mu}$, where $y_{\mu}$ is an integer that labels the corner of the cube, and $\eta_{\mu}=0,1$ labels the sites within a cube.
Using the identity $\tr{(\Gamma^{\dagger}_{\eta}\Gamma_{\eta'} + B^{\dagger}_{\eta}B_{\eta'})} = 4 \delta_{\eta \eta'}$, one can invert the relation in (\ref{SpinTasteTransfo2+1a}) and (\ref{SpinTasteTransfo2+1b}) to obtain
\beq
\label{InverseSpinTasteTransfo2+1}
\chi_{\eta}(y) &=& \sqrt{2} \sum_{\alpha, a}(\Gamma^{* \alpha a}_{\eta} u^{\alpha a}(y) + B^{* \alpha a}_{\eta} d^{\alpha a}(y)), \\ \nn
\chib_{\eta}(y) &=& \sqrt{2} \sum_{\alpha, a}(\bar{u}^{\alpha a}(y)\Gamma^{\alpha a}_{\eta} + \bar{d}^{\alpha a}(y)B^{\alpha a}_{\eta}).
\eeq
One can then rewrite the action in the spin-taste basis. For example, the mass term becomes
\beq
&& a^3m\sum_{y,\eta} \chib_{\eta}(y) \chi_{\eta}(y) = \\ \nn && (2a)^3 \sum_y \left( \bar{u}(y)(\bm 1 \otimes \bm 1)u(y) + \bar{d}(y)(\bm 1 \otimes \bm 1)d(y) \right),
\eeq
where we have used the following identities
\beq
&& \sum_{\eta} \Gamma^{\alpha a}_{\eta} \Gamma^{*\beta b}_{\eta} = \sum_{\eta} B^{\alpha a}_{\eta} B^{*\beta b}_{\eta} = 4 \delta_{\alpha \beta} \delta_{a b}, \\
&& \sum_{\eta} \Gamma^{\alpha a}_{\eta} B^{*\beta b}_{\eta} = 0.
\eeq
To rewrite the kinetic term, one first expresses the shifted field as 
\beq
&&\chi_{\eta + \hat{\mu}}(y) = \\  \nn && \delta_{\eta_{\mu},0} \eta_{\mu}(\eta)\sqrt{2} \tr\bigg(\Gamma^{\dagger}_{\eta}\gamma_{\mu} u(y) + B^{\dagger}_{\eta}\beta_{\mu}d(y) \bigg) \\
\nn && + \delta_{\eta_{\mu},1} \eta_{\mu}(\eta)\sqrt{2} \tr\bigg(\Gamma^{\dagger}_{\eta} \gamma_{\mu}u(y+\hat{\mu}) + B^{\dagger}_{\eta}\beta_{\mu}d(y+\hat{\mu}) \bigg).
\eeq
A similar expression exists for the backward shifted field, $\chi_{\eta - \hat{\mu}}(y)$, and thus one arrives at the following form for the staggered action
\beq
\nn
&& S_{st} = \\ \nn && (2a)^3 \sum_{y, \mu} \bigg\{ \bar{u}(y)(\sigma_{\mu} \otimes \bm 1)\partial_{\mu}u(y) \\ \nn && +~\bar{d}(y)(\beta_{\mu} \otimes \bm 1)\partial_{\mu}d(y) 
+  a[\bar{u}(y)(\bm 1 \otimes \sigma^{T}_{\mu}) \partial^2_{\mu}d(y) \\ \nn && +~\bar{d}(y)(\bm 1 \otimes \beta^{T}_{\mu}) \partial^2_{\mu}u(y)] \bigg\} + (2a)^3 m \sum_y [\bar{u}(y)(\bm 1 \otimes \bm 1)u(y) \\  && +~\bar{d}(y)(\bm 1 \otimes \bm 1)d(y)],
\eeq
where $\sigma^{T}_{\mu}$ refers to the transpose. The derivative operator now acts on a lattice of spacing $2a$ and is defined as
\beq
\partial_{\mu} u(y) \equiv \frac{1}{2(2a)} \left( u(y + \hat{\mu}) - u(y - \hat{\mu}) \right).
\eeq
 One can now define a four-component Dirac spinor as follows
\beq
\label{SpinTastePsi}
\Psi(y) = \left( \begin{array}{c} u(y) \\ d(y) \end{array} \right).
\eeq
Using the reducible set of gamma matrices in (\ref{Gammas1}) and (\ref{Gammas2}), one can write the action in the following compact form
\beq
\label{SpinTasteAction2+1}
S_{st} &=& (2a)^3 \sum_{y, \mu} \bigg\{ \bar{\Psi}(y)(\gamma_{\mu} \otimes \bm 1)\partial_{\mu}\Psi(y) \\ \nn &+& a\bar{\Psi}(y)(\tilde{\gamma}_5 \otimes \sigma^{T}_{\mu}) \partial^2_{\mu}\Psi(y) \bigg\}
\\ \nn &+& (2a)^3 m \sum_y \bar{\Psi}(y)(\bm 1 \otimes \bm 1)\Psi(y),
\eeq
Where the second derivative operator is defined as 
\beq
\nn
\partial^2_{\mu} u(y) \equiv \frac{1}{4(2a)^2} ( u(y+2\hat{\mu}) &+& u(y-2\hat{\mu}) \\  &-& 2u(y) ).
\eeq
One sees from (\ref{SpinTasteAction2+1}) that the second derivative term, which is suppressed by a factor of the lattice spacing, is not invariant under a rotation in taste space.

The residual symmetry of the staggered lattice action is $U(1) \times U(1)_{\epsilon}$. The form of these symmetries on the one-component fields is as follows
\beq
\nn
&& \chi(x) \to \exp(i\alpha) \chi(x), \\  &&\bar{\chi}(x) \to \bar{\chi}(x) \exp(-i\alpha), \\ \nn
&& \chi(x) \to \exp(i\beta \epsilon(x)) \chi(x), \\ && \bar{\chi}(x) \to \bar{\chi}(x) \exp(i\beta \epsilon(x)),
\eeq
where $\epsilon(x) \equiv (-1)^{x_0 + x_1 + x_2}$. In terms of the fields $u$ and $d$, these transformations become
\beq
\nn
&&\left(\begin{array}{c} u \\d \end{array}\right) \to \exp(i\alpha) \left(\begin{array}{c} u \\d \end{array}\right), \\ && \left(\begin{array}{cc} \bar{u} & \bar{d} \end{array}\right) \to
 \left(\begin{array}{cc} \bar{u} & \bar{d} \end{array}\right) \exp(-i\alpha), \\
\label{U1}
\nn
&& \left(\begin{array}{c} u \\d \end{array}\right) \to \left(\begin{array}{cc} \cos(\beta) & i\sin(\beta) \\ i\sin(\beta) & \cos(\beta) \end{array}\right) \left(\begin{array}{c} u \\d \end{array}\right), \\ 
&& \left(\begin{array}{cc} \bar{u} & \bar{d} \end{array}\right) \to  \left(\begin{array}{cc} \bar{u} & \bar{d} \end{array}\right) \left(\begin{array}{cc} \cos(\beta) & i\sin(\beta) \\ i\sin(\beta) & \cos(\beta) \end{array}\right).
\eeq
Thus, one can see that the formation of the condensate $\vev{\chib \chi}$ spontaneously breaks the $U(1)_{\epsilon}$ symmetry and leads to the appearance of a single NG boson.

\section{\label{sec:FermionAppendix}Fermion Bilinears}
In our study we are looking for spontaneous symmetry breaking in the presence of an external magnetic field. Typically, this will involve the breaking of the $SU(2)_{\sigma}$ symmetry, where
$SU(2)_{\sigma}$ is the largest non-abelian subgroup of $U(2)_{\sigma}$, described in (\ref{U2Generators}). In this appendix, we list the expressions for the various bilinear operators in terms of the degrees of freedom on the hexagonal lattice
as well as their representation in terms of staggered lattice fermions.

We first introduce the Dirac mass term
\beq
\label{DiracMass}
\tilde{\Delta}_{\sigma} \Bpsi P_{\sigma} \Psi = \tilde{\Delta}_{\sigma} \Psi^{\dagger} \left( \gamma_0  \otimes P_{\sigma} \right) \Psi,
\eeq
which is a triplet with respect to spin and breaks $SU(2)_{\sigma}$ down to $U(1)_{\sigma}$ with the generator $\tilde{\gamma}_{4,5} \otimes P_{\sigma}$. The corresponding order parameter for this term is $\vev{\Bpsi P_{\sigma} \Psi}$, and, written in terms
of Bloch components it can be expressed as
\beq
\label{DiracMassComponents}
\nn
\tilde{\Delta}_{\sigma}: &~&\psi^{\dagger}_{K_+ A \sigma} \psi_{K_+ A \sigma} - \psi^{\dagger}_{K_+ B \sigma}\psi_{K_+ B \sigma} \\
&+& \psi^{\dagger}_{K_- A \sigma}\psi_{K_- A \sigma} - \psi^{\dagger}_{K_- B \sigma} \psi_{K_- B \sigma}.
\eeq
One can interpret a nonzero value for this order parameter as an imbalance of charge between the two sublattices, A and B, corresponding to a charge density wave (CDW).

The Haldane mass term is given by
\beq
\label{HaldaneMass}
\Delta_{\sigma} \Bpsi \left( \tilde{\gamma}_{4,5} \otimes P_{\sigma} \right) \Psi = \Delta_{\sigma} \Psi^{\dagger} \left( \gamma_0 \tilde{\gamma}_{4,5} \otimes P_{\sigma} \right) \Psi,
\eeq
which is a singlet with respect to spin but is odd under time-reversal. The order parameter for the Haldane mass is $\vev{\Bpsi \left( \tilde{\gamma}_{4,5} \otimes P_{\sigma} \right) \Psi}$, and its expression 
in terms of Bloch components is given by
\beq
\label{HaldaneMassComponents}
\nn
\Delta_{\sigma}: &~ & \psi^{\dagger}_{K_+ A \sigma} \psi_{K_+ A \sigma} - \psi^{\dagger}_{K_- A \sigma}\psi_{K_- A \sigma} \\
&-& \psi^{\dagger}_{K_+ B \sigma}\psi_{K_+ B \sigma} + \psi^{\dagger}_{K_- B \sigma} \psi_{K_- B \sigma}.
\eeq
Thus, this order parameter can be seen to represent a charge imbalance between the two valleys, $K_+$ and $K_-$.
To discuss the properties of (\ref{HaldaneMass}) under time-reversal, we introduce the Hamiltonian of the low-energy theory
\beq
\label{EFTHamiltonian}
\mathcal{H}(\vec{k}) = \hbar v_F \tau_0 \otimes \vec{\sigma} \cdot \vec{k},
\eeq
where $\tau_0$ is a two-dimensional unit matrix in valley space which is tensored with the two-dimensional sublattice space. 
Time-reversal invariance imposes the following restriction on the states as well as the valley Hamiltonians themselves
\beq
\label{TReversalStates}
T \psi_{K_+ (A,B)} &=& \psi^{*}_{K_+ (A,B)} = \psi_{K_- (A,B)}, \\
\label{TReversalHamiltonians}
T\mathcal{H}_{K_+} T^{-1} &=& \mathcal{H}^{*}_{K_-}.
\eeq
The relation in (\ref{TReversalStates}) can be shown by inspecting the explicit form of the valley Hamiltonian eigenstates
\beq
\psi^{K_+}_{e.h.} = \frac{1}{\sqrt{2}} \left(\begin{array}{cc} e^{-i\phi_{\vec{k}}/2}  \\ \pm  e^{i\phi_{\vec{k}}/2} \end{array}\right), \\ 
\psi^{K_-}_{e.h.} = \frac{1}{\sqrt{2}} \left(\begin{array}{cc} e^{i\phi_{\vec{k}}/2}  \\ \pm  e^{-i\phi_{\vec{k}}/2} \end{array}\right), 
\eeq
where the $\pm$ refers to electron and hole states respectively, and $\phi_{\vec{k}} \equiv \tan^{-1}\left(\frac{k_y}{k_x}\right)$ describes the orientation of the vector $\vec{k}$ in the plane. 
Applying the transformation in (\ref{TReversalHamiltonians}) to (\ref{HaldaneMass}), one can verify that this term is odd under time-reversal.

On the lattice, we must calculate the relevant condensates in the language of staggered lattice fermions. In $(2+1)$ dimensions, a bilinear in the spin-taste basis takes the form
\beq
\label{SpinTasteBilinear}
\Bpsi(y)\left( \Gamma_S \otimes \Gamma_T \right) \Psi(y),
\eeq
where the four-component Dirac spinor $\Psi$ is defined in (\ref{DiracSpinorBasis}). From the above discussion, one can see that the four-dimensional spin space can be mapped to the 
four-dimensional sublattice-valley subspace and the two-dimensional taste space can be mapped to the two-dimensional spin of the electron. 

The Dirac mass is the usual staggered mass ($\Gamma_S = \Gamma_T = \bm 1$), and takes the form 
\beq
\label{Condensate}
\nn
\Bpsi(y) \left( \bm 1 \otimes \bm 1 \right) \Psi(y) &=& \bar{u}(y) \left( \bm 1 \otimes \bm 1 \right) u(y) \\ \nn
 &+& \bar{d}(y) \left( \bm 1 \otimes \bm 1 \right) d(y) \\
&=& \frac{1}{8} \sum_{\eta} \chib_{\eta}(y) \chi_{\eta}(y),
\eeq
where we have used the following identity 
\beq
\tr{(\Gamma^{\dagger}_{\eta}\Gamma_{\eta'} + B^{\dagger}_{\eta}B_{\eta'})} = 4 \delta_{\eta \eta'}.
\eeq 
We note that in our lattice calculations, we do not isolate the 
individual spin contributions to the condensates ($\Gamma_T = \bm 1$ i.e. $P_{\sigma} \to \bm 1$). 

The Haldane mass in the spin-taste basis takes the following form
\beq
\label{HaldaneSpinTaste}
\Bpsi(y) \left( \tilde{\gamma}_{4,5} \otimes \bm 1 \right) \Psi(y) &=& \bar{u}(y) \left( \bm 1 \otimes \bm 1 \right) u(y) \\ \nn
&-& \bar{d}(y) \left( \bm 1 \otimes \bm 1 \right) d(y), \\
\label{HaldaneOneComponent}
&=& \frac{i}{8} \sum_{\eta_{\mu} \neq \eta'_{\mu}} \bar{\chi}_{\eta}(y) \chi_{\eta'}(y),
\eeq
where we have employed the identity
\beq
\tr \left[ \Gamma^{\dagger}_{\eta}\Gamma_{\eta'} - B^{\dagger}_{\eta}B_{\eta'} \right] = \left\{ \begin{array}{ll} 4i, &  \eta_{\mu} \neq \eta'_{\mu}, \forall \mu \\
                                                                                             0, & \mbox{otherwise}
                                                                                            \end{array} \right..
\eeq
Thus the Haldane mass involves a bilinear with the fields residing on diagonally opposite sites within the cube. This operator is left invariant by the staggered lattice action's $U(1)_{\epsilon}$ symmetry. 

The subject of time-reversal for staggered fermions in $(2+1)$ dimensions also deserves a discussion. In terms of the underlying graphene lattice, time-reversal interchanges Dirac points, reverses spin, and leaves the sublattice degree of freedom intact. Ignoring spin for the moment and using the basis defined in (\ref{DiracSpinorBasis}), the action of time-reversal in the four-dimensional sublattice-valley subspace is as follows
\beq
\label{TRDiracPoint}
\left( \begin{array}{c}  \psi_{A \sigma}(\vec{K}_+ + \vec{p}) \\  \psi_{B \sigma}(\vec{K}_+ + \vec{p})  \\ \psi_{B \sigma}(\vec{K}_- + \vec{p}) \\ \psi_{A \sigma}(\vec{K}_- + \vec{p})  \end{array} \right)   &\to & \left( \begin{array}{c}  \psi_{A \sigma}(\vec{K}_- - \vec{p}) \\  \psi_{B \sigma}(\vec{K}_- - \vec{p})  \\ \psi_{B \sigma}(\vec{K}_+ - \vec{p}) \\ \psi_{A \sigma}(\vec{K}_+ - \vec{p})  \end{array} \right) \\
&=& \gamma_1 \tilde{\gamma}_5 \Psi_{\sigma}(-\vec{p}),
\eeq
where $\psi_{K_{\pm}, A/B, \sigma}(\vec{p}) \equiv \psi_{A/B, \sigma}(\vec{K}_{\pm} + \vec{p})$, and we have written the spinor in momentum space. Time-reversal acts on the spin degree of freedom in the following way
\beq
\label{TRSpin}
\left( \begin{array}{c} \psi_{K_{\pm}, A/B, \uparrow} \\ \psi_{K_{\pm}, A/B, \downarrow} \end{array} \right) &\to& \left( \begin{array}{c} \psi_{K_{\pm}, A/B, \downarrow} \\ -\psi_{K_{\pm}, A/B, \uparrow} \end{array} \right), \\
&=& i\sigma_2 \left( \begin{array}{c} \psi_{K_{\pm}, A/B, \uparrow} \\ \psi_{K_{\pm}, A/B, \downarrow} \end{array} \right). 
\eeq
Combining the action of time-reversal on the four-dimensional sublattice-valley space, which composes the four-dimensional spinor structure, and on the two-dimensional spin space, which composes the two-dimensional ``flavor" space, we obtain the following result in momentum space
\beq
\Psi(\vec{p}) \to \left( \gamma_1 \tilde{\gamma}_5  \otimes i\sigma_2 \right) \Psi(-\vec{p}). 
\eeq
For the coordinate space spinor, time-reversal takes the following form
\beq
\Psi(\vec{x},  t) \to \left( \gamma_1 \tilde{\gamma}_5  \otimes i\sigma_2 \right) \Psi(\vec{x}, -t).
\eeq
One can then show that the continuum action in Minkowski space is left invariant by the following transformations \cite{Gusynin}
\beq
\label{TRPsi}
\Psi(\vec{x},t) &\to & \left( \gamma_1 \tilde{\gamma}_5 \otimes i\sigma_2 \right) \Psi(\vec{x}, -t), \\
\label{TRBpsi}
\Bpsi(\vec{x},t) &\to & - \Bpsi(\vec{x}, -t) \left( \gamma_1 \tilde{\gamma}_5 \otimes i\sigma_2 \right) \\
\label{TRA_0}
A_0(\vec{x}, t) &\to & -A_0(\vec{x}, -t),
\eeq
where, for a fermion bilinear of the form $\Bpsi A \Psi$, due to the fact that time-reversal, $\mathcal{T}$, is an anti-unitary operator, one has $\mathcal{T} A \mathcal{T}^{-1} = A^*$.
In Euclidean space, the time-reversal transformation takes a different form. In this case, time is not distinguished from the spatial coordinates by a relative minus sign in the metric. In $(2+1)$ dimensions, the time-reversal transformation on staggered fermions in the spin-taste basis is as follows
\beq
\nn
\Psi(\vec{y}, y_0) &\to& -i\tilde{\gamma}_5 \ \gamma_1 \gamma_2 \Psi(\vec{y}, -y_0), \\ \nn
&=&  \left( \begin{array}{cc} 0 & \sigma_0 \\ -\sigma_0 & 0 \end{array} \right) \left( \begin{array}{c} u(\vec{y}, -y_0) \\ d(\vec{y}, -y_0) \end{array} \right), \\
\label{TRSpinTaste}
&=&  \left( \begin{array}{c} \sigma_0  d(\vec{y}, -y_0) \\ -\sigma_0 u(\vec{y}, -y_0) \end{array} \right).
\eeq
To see the effect of time-reversal on the one-component basis, we note the following identity
\beq
\label{OneComponentProjection}
\chi_{\eta}(y) = \sqrt{2} \tr \left\{ \left(\Gamma^{\dagger}_{\eta},~ B^{\dagger}_{\eta} \right) \left( \begin{array}{c} u(y) \\ d(y) \end{array} \right) \right\},
\eeq
where we regard the spinor as a matrix with the Dirac index representing the row and the taste index representing the column. Thus to find the action of time-reversal on the one-component spinor we must evaluate the following expression
\beq
\tr \left\{ \left(\Gamma^{\dagger}_{\eta},~ B^{\dagger}_{\eta} \right) \left( \begin{array}{c} \sigma_0  d(\vec{y},-y_0) \\ -\sigma_0 u(\vec{y},-y_0) \end{array} \right) \right\},
\eeq
which represents the projection of the spin-taste basis onto the one-component basis. Using the expressions for the spinors $u$ and $d$ in (\ref{SpinTasteTransfo2+1a}) and (\ref{SpinTasteTransfo2+1b}), one must compute $\tr \left(  \Gamma^{\dagger}_{\eta} \sigma_0 B_{\eta'} - B^{\dagger}_{\eta} \sigma_0 \Gamma_{\eta'} \right)$.
The only nonzero contribution comes when $\eta' = \tilde{\eta} \equiv \left( \eta_0 \pm 1, \eta_1, \eta_2 \right)$, where the ``$+$" corresponds to $\eta_0=0$ and the ``$-$" corresponds to $\eta_0=1$. The result is thus
\beq
\nn
&& \sqrt{2} \tr \left\{ \left(\Gamma^{\dagger}_{\eta},~ B^{\dagger}_{\eta} \right) \left( \begin{array}{c} \sigma_0  d(\vec{y},-y_0) \\ -\sigma_0 u(\vec{y},-y_0) \end{array} \right) \right\} \\  && = (-)^{\eta_0 + \eta_1 + \eta_2} \chi_{\tilde{\eta}}(\vec{y}, -y_0).
\eeq
This implies that time-reversal induces the following transformation on the one-component staggered field
\beq
\label{TROneComponent}
\chi_{\eta}(\vec{y}, y_0) \to (-)^{\eta_0 + \eta_1 +  \eta_2}\chi_{\tilde{\eta}}(\vec{y}, -y_0),
\eeq
where an analogous expression holds for $\chib$. Using the expressions for the time-reversal transformation in the spin-taste as well as the one-compent basis, one can show that the operators in (\ref{HaldaneSpinTaste}) and (\ref{HaldaneOneComponent}) change sign and are thus, indeed, time-reversal-odd.


\begin{thebibliography}{99}

\bibitem{Novoselov}
K.~S.~Novoselov \etal, Science {\bf 306}, 666, (2004), [arXiv:cond-mat/0410550 [cond-mat.mtrl-sci]]

\bibitem{CastroNeto}
A.~H.~Castro Neto \etal, Rev.\ Mod.\ Phys.\ {\bf 81}, 109 (2009), [arXiv:0709.1163 [cond-mat.other]]

\bibitem{Goerbig}
M.~O.~Goerbig, Rev.\ Mod.\ Phys.\ {\bf 83}, 1193  (2011), [arXiv:1004.3396 [cond-mat.mes-hall]]

\bibitem{Drut1}
J.~E.~Drut and T.~A.~Lahde, Phys.\ Rev.\ Lett.\ {\bf 102}, 026802 (2009), [arXiv:0807.0834 [cond-mat.str-el]]

\bibitem{Drut2}
J.~E.~Drut and T.~A.~Lahde, Phys.\ Rev.\ B{\bf79}, 165425, (2009), [arXiv:0901.0584 [cond-mat.str-el]]

\bibitem{Hands1}
W.~Armour, S.~Hands, and C.~Strouthos, Phys.\ Rev.\ B{\bf 81}, 125105 (2010), [arXiv:0910.5646 [cond-mat.str-el]]

\bibitem{Hands2}
W.~Armour, S.~Hands, and C.~Strouthos, Phys.\ Rev.\ B{\bf 84}, 075123 (2011), [arXiv:1105.1043 [cond-mat.str-el]]

\bibitem{ZhangQHE}
Y.~Zhang \etal, Phys.\ Rev.\ Lett.\ {\bf 96}, 136806 (2006), [arXiv:cond-mat/0602649 [cond-mat.mes-hall]]

\bibitem{JiangQHE}
Z.~Jiang, Y.~Zhang, H.~L.~Stormer, and P.~Kim, Phys.\ Rev.\ Lett.\ {\bf 99}, 106802 (2007), [arXiv:0705.1102 [cond-mat.mes-hall]]

\bibitem{Kennett}
B.~Roy, M.~P.~Kennett, and S.~Das Sarma, Phys.\ Rev.\ B{\bf 90}, 201409, [arXiv:1406.5184]

\bibitem{Buividovich}
P.~Buividovich, M.~Chernodub, E.~V.~Luschevskaya, and M.~I.~Polikarpov, Phys.\ Lett.\ B {\bf 682}, 484 (2010), [arXiv:0812.1740 [hep-lat]]

\bibitem{Braguta}
V.~V.~Braguta, P.~V.~Buividovich, T.~Kalaydzhyan, S.~V.~Kuznetsov, and M.~I.~Polikarpov, PoS LATTICE2010, 190 (2010), [arXiv:1011.3795 [hep-lat]]

\bibitem{Cohen}
T.~D.~Cohen, D.~A.~McGady, and E.~S.~Werbos, Phys.\ Rev.\ C {\bf 76}, 055201 (2007), [arXiv:0706.3208 [hep-ph]]

\bibitem{Bali1}
G.~S.~Bali, F.~Bruckmann, G.~Endrodi, Z.~Fodor, and S.~D.~Katz, PoS LATTICE2011, 192 (2011), [arXiv:1111.5155 [hep-lat]]


\bibitem{Bali2}
G.~S.~Bali, F.~Bruckmann, G.~Endrodi, Z.~Fodor, and S.~D.~Katz, J.\ High Energy Phys.\ {\bf 2012}, 044 (2012), [arXiv:1111.4956 [hep-lat]]

\bibitem{DPF2015}
C.~DeTar, C.~Winterowd, and S.~Zafeiropoulos, [arXiv:1509.06432 [hep-lat]]

\bibitem{GrapheneLetter}
C.~DeTar, C.~Winterowd, and S.~Zafeiropoulos, [arXiv:1607.03137 [hep-lat]]

\bibitem{DrutSon}
J.~E.~Drut and D.~T.~Son, Phys.\ Rev.\ B{\bf 77}, 075115 (2008), [arXiv:0710.1315 [cond-mat.str-el]]

\bibitem{Aleiner}
I.~L.~Aleiner, D.~E.~Kharzeev, and A.~M.~Tsvelik, Phys.\ Rev.\ B{\bf 76}, 195415 (2007), [arXiv:0708.0394 [cond-mat.mes-hall]]

\bibitem{Kharitonov}
M.~Kharitonov, Phys.\ Rev.\ B{\bf 85}, 155439 (2012), [arXiv:1103.6285 [cond-mat.str-el]]

\bibitem{DeGrandDeTar}
T.~DeGrand and C.~DeTar, {\it Lattice Methods for Quantum Chromodynamics}, World Scientific (2006)

\bibitem{KogutStrouthos}
J.~B.~Kogut and C.~G.~Strouthos, Phys.\ Rev.\ D{\bf 71}, 094012 (2005), [arXiv:hep-lat/0501003]

\bibitem{NielsenNinomiya}
H.~B.~Nielsen and M.~Ninomiya, Nucl.\ Phys.\ {\bf 185}, 20 (1981)

\bibitem{KogutSusskind}
J.~B.~Kogut and L.~Susskind, Phys.\ Rev.\ D{\bf 9}, 3501 (1974)

\bibitem{Orginos}
K.~Orginos, R.~Sugar, and D.~Toussaint, Phys.\ Rev.\ D{\bf 60}, 054503 (1999), [arXiv:hep-lat/9903032]

\bibitem{MILCStaggeredReview}
A.~Bazavov \etal, Rev.\ Mod.\ Phys.\ {\bf 82}, 1349 (2010), [arXiv:0903.3598 [hep-lat]]

\bibitem{LagaeSinclair}
J.-F.~Laga\"e and D.~K.~Sinclair, Nucl.\ Phys.\ Proc.\ Suppl.\ {\bf 63}, 892 (1998), [arXiv:hep-lat/9709035]

\bibitem{LepageMackenzie}
G.~P.~Lepage and P.~B.~Mackenzie, Phys.\ Rev.\ D{\bf 48}, 2250 (1993), [arXiv:hep-lat/9209022]

\bibitem{Naik}
S.~Naik, Nucl.\ Phys.\ B{\bf 316}, 238 (1989)

\bibitem{Giedt}
J.~Giedt, A.~Skinner, and S.~Nayak, Phys.\ Rev.\ B{\bf 83}, 045420 (2011), [arXiv:0911.4316 [cond-mat.str-el]]

\bibitem{Drut3}
J.~E.~Drut, T.~A.~Lahde, and L.~Suoranta, [arXiv:1002,1273], (2010), [arXiv:1002.1273 [cond-mat.str-el]]

\bibitem{WieseAlHashimi}
M.~H.~Al-Hashimi and U.~J.~Wiese, Annals Phys.\ {\bf 324}, 343 (2009), [arXiv:0807.0630 [quant-ph]]

\bibitem{Miransky1}
V.~Gusynin, V.~Miransky, and I.~Shovkovy, Phys.\ Rev.\ Lett.\ {\bf 73}, 3499 (1994), [arXiv:hep-ph/9405262]

\bibitem{Miransky2}
V.~Gusynin, V.~Miransky, and I.~Shovkovy, Phys.\ Lett.\ B{\bf 349}, 477 (1995), [arXiv:hep-ph/9412257]

\bibitem{Miransky3}
V.~Gusynin, V.~Miransky, and I.~Shovkovy, Phys.\ Rev.\ D{\bf 52}, 4718 (1995), [arXiv:hep-ph/9501304]

\bibitem{Miransky4}
V.~Gusynin, V.~Miransky, and I.~Shovkovy, Nucl.\ Phys.\ B{\bf 462}, 249 (1996), [arXiv:hep-ph/9509320]

\bibitem{Khveshchenko}
D.~V.~Khveshchenko, Phys.\ Rev.\ Lett.\ {\bf 87}, 206401 (2001), [arXiv:cond-mat/0106261]

\bibitem{MiranskyGraphene1}
E.~Gorbar, V.~Gusynin, V.~Miransky, and I.~Shovkovy, Phys.\ Rev.\ B{\bf 66}, 045108 (2002), [arXiv:cond-mat/0202422]

\bibitem{MiranskyGraphene2}
E.~Gorbar, V.~Gusynin, V.~Miransky, and I.~Shovkovy, Phys.\ Rev.\ B{\bf 78}, 085437 (2008), [arXiv:0806.0846 [cond-mat.mes-hall]]

\bibitem{MiranskyGraphene3}
E.~Gorbar, V.~Gusynin, V.~Miransky, and I.~Shovkovy, Phys.\ Scripta T{\bf 146}, 014018 (2012), [arXiv:1105.1360 [cond-mat.mes-hall]]

\bibitem{Shovkovy}
I.~Shovkovy, {\it Lect. Notes Phys.} {\bf 871}, 13 (2013), [arXiv:1207.5081 [hep-ph]]

\bibitem{Schwinger}
J.~S.~Schwinger, Phys.\ Rev.\ {\bf 82}, 664 (1951)

\bibitem{DittrichGies}
W.~Dittrich and H.~Gies, Phys.\ Lett.\ B{\bf 392}, 182 (1997), [arXiv:hep-th/9609197]

\bibitem{DittrichReuter}
W.~Dittrich and M.~Reuter, {\it Effective Lagrangians in Quantum Electrodynamics}, Springer (1985)




\bibitem{Haldane}
F.~Haldane, Phys.\ Rev.\ Lett.\ {\bf  61}, 2015 (1988)

\bibitem{PhiAlgorithm}
S.~A.~Gottlieb, W.~Liu, D.~Toussaint, R.~L.~Renken, and R.~L.~Sugar, Phys.\ Rev.\ D{\bf 35}, 2531 (1987)

\bibitem{KogutDuane}
S.~Duane and J.~B.~Kogut, Phys.\ Rev.\ Lett.\ {\bf 55}, 2774 (1985)

\bibitem{Polikarpov}
D.~L.~Boyda, V.~V.~Braguta, S.~N.~Valgushev, M.~I.~Polikarpov, and M.~V.~Ulybyshev, Phys.\ Rev.\ B{\bf 89}, 245404 (2014), [arXiv:1308.2814 [hep-lat]]

\bibitem{Cosmai}
P.~Cea, L.~Cosmai, P.~Giudice, A.~Papa, Phys.\ Rev.\ D{\bf85}, 094505 (2012), [arXiv:1204.6112 [hep-lat]]

\bibitem{Yang}
K.~Yang, [arXiv:cond-mat/0703757]

\bibitem{GonzalezHaldaneMass}
J.~Gonzalez, JHEP {\bf 07}, 175 (2013), 	[arXiv:1211.3905 [cond-mat.mes-hall]]

\bibitem{SmilgaShushpanov}
I.~A.~Shushpanov and A.~V.~Smilga, Phys.\ Lett.\ B{\bf 402}, 351 (1997), [arXiv:hep-ph/9703201]

\bibitem{Burkitt}
C.~Burden and A.~N.~Burkitt, Eur.\ Phys.\ Lett.\ {\bf 3}, 545 (1987)

\bibitem{Gusynin}
V.~P.~Gusynin, S.~G.~Sharapov, and J.~P.~Carbotte, Int.\ J.\ Mod.\ Phys.\ B{\bf 21}, 4611 (2008) 

\end{thebibliography}
\end{document}